\documentclass[preprint]{aastex6}

\usepackage{amsmath,amssymb}
\usepackage{graphicx}
\usepackage{hyperref}
\DeclareMathOperator{\sign}{sign}

\slugcomment{Submit for publication in ApJS}
\shorttitle{MPI-AMRVAC 2.0}
\shortauthors{Xia et al.}

\begin{document}

\title{\texttt{MPI-AMRVAC 2.0} for Solar and Astrophysical Applications}

\author{C. Xia\altaffilmark{1}, J. Teunissen\altaffilmark{1}, 
I. El Mellah\altaffilmark{1}, E. Chan\'e\altaffilmark{1}, R. Keppens\altaffilmark{1}}
\altaffiltext{1}{Centre for mathematical Plasma Astrophysics, Department of
Mathematics, KU Leuven, Celestijnenlaan 200B, 3001 Leuven, Belgium}

\begin{abstract}
We report on the development of \texttt{MPI-AMRVAC} version 2.0, which is an
open-source framework for parallel, grid-adaptive simulations of hydrodynamic
and magnetohydrodynamic (MHD) astrophysical applications. The framework now
supports radial grid stretching in combination with adaptive mesh refinement (AMR).
The advantages of this combined approach are demonstrated with
one-dimensional, two-dimensional and three-dimensional examples of spherically
symmetric Bondi accretion, steady planar Bondi-Hoyle-Lyttleton flows, and wind
accretion in Supergiant X-ray binaries. Another improvement is support for the
generic splitting of any background magnetic field. We present several tests
relevant for solar physics applications to demonstrate the advantages of field
splitting on accuracy and robustness in extremely low plasma $\beta$
environments: a static magnetic flux rope, a magnetic null-point, and magnetic
reconnection in a current sheet with either uniform or anomalous resistivity.
Our implementation for treating anisotropic thermal conduction in
multi-dimensional MHD applications is also described, which generalizes the
original slope limited symmetric scheme from 2D to 3D. We perform ring
diffusion tests that demonstrate its accuracy and robustness, and show that it
prevents the unphysical thermal flux present in traditional schemes. The
improved parallel scaling of the code is demonstrated with 3D AMR simulations
of solar coronal rain, which show satisfactory strong scaling up to 2000
cores. Other framework improvements are also reported: the modernization and
reorganization into a library, the handling of automatic regression tests, the
use of inline/online Doxygen documentation, and a new future-proof data format
for input/output.
\end{abstract}


\section{Introduction}\label{intro}

The solar and astrophysical research community can benefit from a whole array of publicly available software frameworks that are suited to model instrinsically multi-scale problems on modern computing infrastructure. A recurring ingredient in coping with the multi-scale nature is their capability of using adaptive mesh refinement (AMR). This is essential, e.g. in coping with Sun-to-Earth modeling of solar coronal mass ejections~\citep{Manchester2004}, in many applications which extend magnetohydrodynamics (MHD) into special relativistic  regimes~\citep{Mignone12,Keppens12}, in applications recreating conditions in laser-plasma interactions~\citep{Tzeferacosetal2015}, in simulations focusing on the formation and early evolution of disk galaxies in magnetized interstellar medium~\citep{WangAbel2009}, or in fact in many problems where self-gravitational collapse is at play and AMR-MHD is combined with Poisson solvers~\citep{Ziegler2005}. AMR-MHD is currently available in a large -- and still growing -- variety of codes (e.g. \texttt{BATS-R-US, Pluto, MPI-AMRVAC, FLASH, Enzo, Nirvana, Ramses, Athena, ASTRO-BEAR,} \ldots), and it drove the development of novel MHD algorithms~\citep[e.g.,][]{Fromangetal2006,Rossmanith2004,Stoneetal2008,Cunninghametal2009}. For example, recent Sun-to-Earth model efforts combine the use of a near-Sun overset Yin-Yang grid, with a cartesian AMR grid to handle the further heliosphere dynamics~\citep{Feng11}, where the AMR part is based on the \texttt{PARAMESH}~\citep{MacNeice00} implementation.

In this paper, we report an upgrade to the open source \texttt{MPI-AMRVAC} software ~\citep{Porth14}, 
which was initially developed as a tool to study hyperbolic partial differential equation (PDE) systems such 
as the Euler system or ideal MHD with patch-based or hybrid-block-based AMR meshes \citep{Keppens03,Holst07}. Using the loop annotation syntax (\texttt{LASY})
\citep{Toth97}, the software is written in Fortran 90 plus Perl annotations
so that the same code can be translated by a Perl preprocessor into 
any dimensional pure Fortran code before compilation. With the \texttt{LASY} implementation, 
typical expressions appearing in multidimensional simulation codes can be written in a much 
more concise form, less prone to mistakes. Adding geometric source terms, 
one-to-three dimensional AMR meshes can be in Cartesian, polar, 
cylindrical, or spherical coordinate systems. As a result, \texttt{MPI-AMRVAC} matured to a multi-physics software framework with modules to
handle special relativistic hydro to MHD problems \citep{Meliani07,Holst08,Keppens12} and with the 
ability to couple different PDE models in localized domain regions across
block-based AMR meshes~\citep{Keppens14}. To ensure efficient scaling on realistic applications, the AMR strategy evolved to a pure quadtree-octree block-based AMR hierarchy~\citep{Keppens12,Porth14}. In~\citet{Porth14}, we reported on conservative fourth-order 
finite difference schemes, and new modules to allow simulating hydrodynamics with dust 
coupling~\citep{hendrix14} or Hall MHD~\citep{Leroy17}.
Selected applications of \texttt{MPI-AMRVAC} focus on shock-dominated special relativistic hydro problems in the context of gamma-ray burst blast waves and their afterglow \citep{Meliani10,Vlasis11}, and up to full three-dimensional (3D) modeling of the precessing jets in the X-ray binary SS433~\citep{Monceau14}. Relativistic MHD applications studied propagation aspects of magnetized relativistic jets~\citep{Keppens08}, to detailed modeling of  the Crab nebula~\citep{Porth14MN,Porth14RT}. Newtonian hydro applications addressed accretion processes onto compact objects~\citep{ElMellah2015}, the wind collision and dust redistribution in the Wolf-Rayet 98a rotating pinwheel nebula~\citep{hendrix16}, and MHD is employed in many
solar challenges, such as modeling the in-situ 
formation of solar prominences \citep{Xia12,Xia14,Xia16}, solar coronal rain 
dynamics \citep{Fang13,Fang15,Xia17}, or the initiation of coronal mass ejections 
\citep{ZhaoX17,Mei17}.

Recently, general relativistic fluid or plasma dynamics in arbitrary spacetimes can be handled by the Black Hole Accretion Code (\texttt{BHAC}), an \texttt{MPI-AMRVAC} spin-off that shares its algorithmic versatility and efficient block-based AMR ability~\citep{Porth17}.  \texttt{MPI-AMRVAC} development focused on more Newtonian physics, with solar physics oriented functionality to extrapolate vector magnetic field distributions into coronal volumes using magnetofrictional methods~\citep{Guo16a,Guo16b}. This augments the earlier potential to linear-force-free field extrapolation in local Cartesian boxes or global 
spherical coordinates \citep{Porth14}, and paves the way to future data-driven solar applications. 
Driven by these ambitions as well as by several recent applications, many new 
functionalities have been developed. Since several of them can be of general interest 
to the astrophysical community and have not been documented elsewhere, we present them here in more detail, with an emphasis on demonstrating their advantages in a series of tests.
We start with hydro to MHD applications where we employ radially stretched meshes in curvilinear coordinates in Section~\ref{sec:stretched}. This is shown to be crucial in accretion flow (or outflow) setups, where we wish to concentrate resolution around the accreting object near the inner radial boundary  
and alleviate the deformation of the cells at large radial distances. For MHD problems, 
the technique of magnetic field splitting, invented to
handle low-$\beta$ shock-dominated plasma, was originally limited to potential (current-free)  background
fields~\citep{Tanaka94}. We report an extension of this technique
to allow any magnetic field to be the time-independent split-off part in Section~\ref{sec:b0field}. 
To accurately solve anisotropic thermal conduction in magnetized
plasma, we present an extension to 3D of a slope-limited symmetric scheme in Section \ref{sec:conduction}, which is shown to handle sharp temperature gradients very well. We report on scaling tests where these methods are adopted in full 3D coronal rain applications in Section~\ref{sec:scaling}. Other useful, more generic code framework improvements are reported in Section~\ref{sec:framework-improvements}, which collectively define what we introduce as \texttt{MPI-AMRVAC 2.0}. 

\section{Radially stretched meshes}\label{sec:stretched}

Many physical problems rely on the use of spherical, polar or 
cylindrical meshes, and at the same time intend to monitor the evolution of gas or plasma flows over 
several spatial orders of magnitude. In astrophysical configurations, 
the interplay between gravitation and in- or outflows is a recurring ingredient that would benefit from combining e.g. a spherical mesh with a gradual radial stretching of the cells, in combination with block-adaptivity. Here, we report on one-dimensional (1D) to fully 3D tests or applications where this is exploited.

\subsection{1D Bondi transonic accretion}\label{sec:ex_1}

The simplest hydrodynamical representation of the accretion of matter by a point mass has
 been analysed early on by \citet{Bondi1952}. It is an isotropic problem where only radial 
dependencies and components are considered. It describes 
how a non-zero temperature flow moving at a subsonic speed at infinity towards
a central object of mass $M_*$ can establish a steady, smooth and transonic inflow solution.
This unique continuous, transonic inflow solution $v_r(r)$ does not display any shock. Provided the evolution is adiabatic, it means that the flow is isentropic: the pressure $p(r)$ can be 
deduced from the mass density $\rho(r)$, using the fixed adiabatic index $\gamma$ of the flow ($p\rho^{-\gamma}$ is independent of radius $r$). 
It is convenient to adopt the
following normalization quantities:
\begin{enumerate}
\item the mass density at infinity, $\rho_{\infty}$,
\item the sound speed at infinity, $c_{\infty}$,
\item the radius $R_0=2GM_*/c_{\infty}^2$, where $M_*$ is the mass of the accretor and $G$ 
the gravitational constant.
\end{enumerate}
Using these scalings, the mass per unit time crossing a sphere of a given radius (the mass accretion rate, $\dot{M}$) and the pressure are normalized with $\rho_{\infty}c_{\infty}R_0^2$ and 
$\rho_{\infty}c_{\infty}^2$, respectively. From now on, we work with normalized
dimensionless variables. The analytic expressions below assume the specific case of a zero speed at infinity, but the conclusions below remain qualitatively the same in the general case.

The fundamental equations of this problem are the mass continuity equation 
\begin{eqnarray}
\frac{\partial \rho}{\partial t}+\frac{1}{r^2}\frac{\partial (r^2 \rho v_r)}{\partial r} & = & 0 \,,
\end{eqnarray}
linear 
momentum conservation with gravity as a source term 
 \begin{eqnarray}
\frac{\partial (\rho v_r)}{\partial t}+\frac{1}{r^2}\frac{\partial (r^2 \rho v_r^2 )}{\partial r} +\frac{\partial p}{\partial r}& = &  - \frac{\rho}{2 r^2} \,,
\end{eqnarray}
and the isentropic relation between (scaled)
pressure and density as $p\rho^{-\gamma}=p_\infty/(\rho_\infty c_\infty^2)=1/\gamma$.
The steady-state solution admits a radius at which it becomes supersonic, called the sonic 
radius $r_s$, whose expression can be analytically derived to be 
$r_s={(5-3\gamma)}/{8}$. The isothermal case (i.e. $\gamma=1$) has its associated sonic radius at $r_{s,0}=1/4$ (this is identical for the corresponding unique outflowing Parker isothermal wind solution).  The normalized mass accretion rate can also be 
determined from the adiabatic index:
\begin{equation}
\label{eq:mdot}
\dot{M}=\frac{\pi}{4}\left(\frac{2}{5-3\gamma}\right)^\frac{5-3\gamma}{2\left(\gamma-1\right)} \,.
\end{equation}
For derivations of these results above, in the more general case of a non zero 
subsonic speed at infinity, the reader might refer to Section 4.2 of \citet{ElMellah2016}.
Below $r=1$ (i.e. below the position of the scaling radius $R_0$) the 
properties of the flow significantly depart from their values at infinity. For adiabatic 
indexes close to the maximum value of $5/3$ that it can take, the sonic 
radius $r_s$ becomes very small. We want to have the sonic 
radius within the simulation space, to have a trivial inner boundary treatment where inflow is supersonic. It then becomes advantagous to resort to a stretched mesh.

We work on domains extending from a radius of 0.01 (inner boundary) to 10 (outer 
boundary). We rely on a second order two-step HLLC method~\citep{Toro94}, both for the predictor and the full 
time step, with the Koren slope limiter~\citep{Koren93} used in the limited reconstruction of the primitive variables $\rho$ and $v_r$, to obtain cell-face values. To design our initial conditions and fix the outer $r=10$ inflow boundary conditions, we reformulate the fundamental equations 
to obtain a transcendental equation which indirectly gives the mass density as a function 
of radius $r$
\begin{equation}
\frac{\rho^{\gamma-1}-1}{\gamma-1}-\frac{1}{2r}+\frac{\dot{M}^2}{32\pi^2}\frac{1}{\rho^2r^4}=0\,.
\end{equation}
We numerically solve this relation with a Newton-Raphson method, which behaves 
correctly except very near the sonic radius where the algorithm may yield non monotonous results. In this restricted region, we compute the mass density profile by linearly interpolating between points further away from the sonic radius.  The 
obtained mass density profile is used as initial condition and sets the fixed mass density in the ghost cells beyond $r=10$.  For the linear momentum $\rho v_r$, 
we deduce its radial profile from Eq.~(\ref{eq:mdot}) and similarly implement it both as initial 
condition and as forced outer boundary condition. For the inner boundary conditions, we copy the mass density in the ghost cells and then guarantee the continuity of the mass flux and Bernoulli invariant \citep[as in][]{ElMellah2015}. We then use \texttt{MPI-AMRVAC} to time-advance the solution to
retrieve the stationary analytical solution, on three different kinds of 
radial spherical meshes :
\begin{enumerate}
\item a non stretched grid with a total number of cells $N=16388$, without AMR,
\item a stretched grid with a much reduced total number of cells ($N=132$), stretched such that the relative resolution at 
the inner boundary remains the same as for the previous grid, without AMR, 
\item the same stretched grid as above, but with a total of 5 levels of mesh refinement, where blocks which contain the analytic sonic radius are refined. The refined grid is computed at the initial state and does not evolve afterwhile (static AMR). 
\end{enumerate}

\begin{figure}
\includegraphics[width=\textwidth]{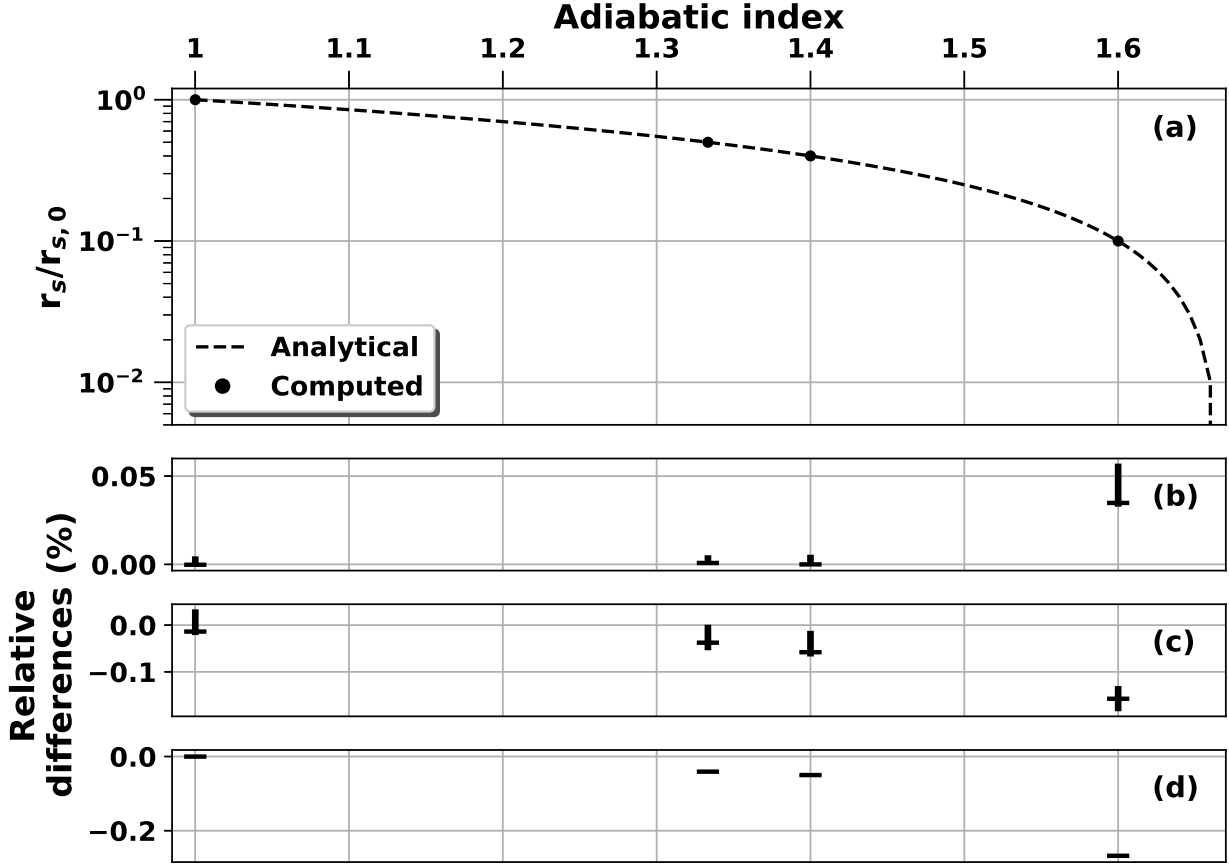}
\caption{(a) Sonic radius $r_s$ relative to its value $r_{s,0}$ at $\gamma=1$ as a 
function of the adiabatic index. The dashed line represents the analytic solution given 
by $r_s={(5-3\gamma)}/{8}$, while the 4 dots are the sonic radii measured from the steady-state solution 
computed by \texttt{MPI-AMRVAC}. (b,c,d) Relative differences between the analytic and 
computed sonic radii for a non-stretched high resolution grid, a stretched grid and a 
stretched grid with AMR.}
\label{fig:rs_gm}
\end{figure}

Within a few dynamical times, the profiles relax towards a numerical equilibrium 
close to the solution determined with the Newton--Raphson above. To quantify the relaxed steady states, we measure the 
position of the sonic radius obtained for each of the three meshes, and this for 4 different adiabatic indexes. 
The results are summarized in Figure~\ref{fig:rs_gm}. By eye, it is impossible to 
distinguish the numerically obtained sonic radii (dots) from the analytic solution 
(dashed curve) displayed in the upper panel, whatever the mesh. Therefore, we plotted the 
relative differences in the three lower panels for the non stretched grid (b), a stretched 
grid (c) and a stretched grid with AMR (d). The error bars are evaluated by considering 
one hundredth of the spatial resolution in the vicinity of the sonic radius. We can 
conclude from those results that analytic sanity checks such as the position of the sonic 
radius are retrieved within a few tenths of percent with \texttt{MPI-AMRVAC}. 

\begin{figure}
\includegraphics[width=\textwidth]{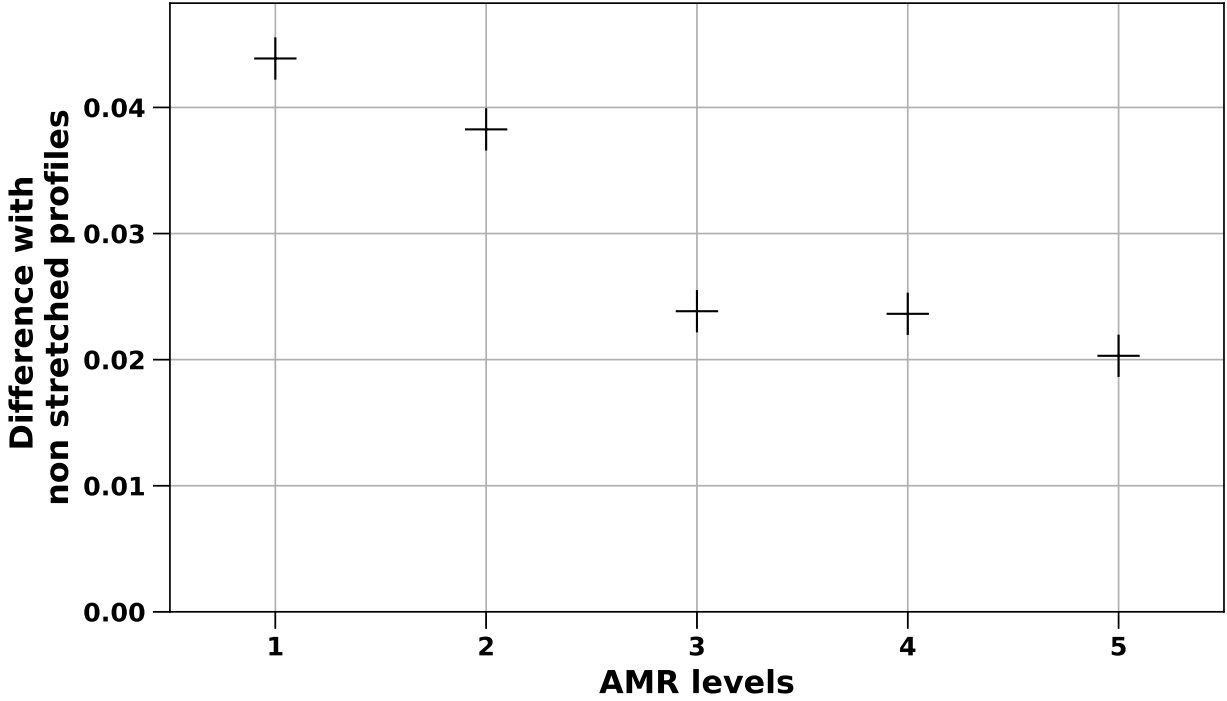}
\caption{Quantification of the departure of the relaxed mass density profiles obtained 
with stretched meshes, for different AMR levels, with respect to the profile obtained 
with a non stretched mesh. Those results were obtained for $\gamma=1.4$.}
\label{fig:match}
\end{figure}

Apart from this local agreement on the sonic point location, we can also quantify an error 
estimate for the entire relaxed mass density profiles obtained. For the case where 
$\gamma=1.4$, we compare runs with different AMR levels employing stretched meshes, to the relaxed mass density profile 
obtained with the non-stretched, high resolution mesh above. We computed the 
$L_2$-norm between data points on stretched, AMR meshes and their counterpart on the non-stretched mesh (linearly 
interpolated to the same radial positions). Figure~\ref{fig:match} shows that 
the matching between the solutions obtained increases as the number of maximum AMR levels 
rises.

\begin{figure}
\includegraphics[width=\textwidth]{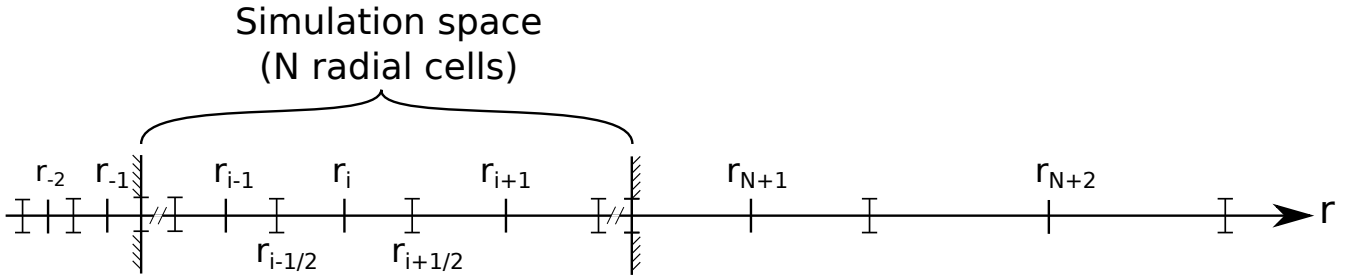}
\caption{Uni-dimensional representation of the radial stretching with the cell centers marked 
with a vertical line and the cell edges with a back-to-back double bracket. The cell centers 
are always mid-way between two consecutive edges, but the reverse is not true. The inner and 
outer edges are marked by longer vertical lines with dashes. There are two ghost cells on 
each side ($n=2$) in this example.}
\label{fig:rule}
\end{figure}

\subsection{Radial stretching principle}

The radial stretching is introduced both to decrease the number of radial cells
and, in two-dimensional (2D) and 3D, to alleviate the deformation of cells when the radius of the inner radial boundary of 
the grid is orders of magnitude smaller than the radius of the outer boundary. To keep a constant 
cell aspect ratio all over a 2D $(r, \theta)$ grid, the radial step size $\Delta r_i$ must obey the following relation :
\begin{equation}
\label{eq:cst_aspect_ratio}
\frac{\Delta r_i}{r_i\Delta\theta}=\zeta
\end{equation}
where $r_i$ is the local radius, $\Delta\theta$ the uniform angular step (the latitudinal one for a 
spherical $(r, \theta, \varphi)$ mesh where $\theta$ is the colatitude) and $\zeta$ is a fixed aspect ratio. Since the radial step is then
proportional to the radius, we obtain a kind of self-similar grid from the innermost to the outermost regions. We set the radial position $r_i$ of 
the center of any cell indexed $i$ mid-way between the two surrounding interfaces, $r_{i-1}$ and 
$r_{i+1}$, see Figure~\ref{fig:rule}. The recursive relationship which builds up the sequence 
of radial positions for the cell centers is given by :
\begin{equation}
r_{i+1}=r_i+\frac{\Delta r_i}{2}+\frac{\Delta r_{i+1}}{2} \,,
\end{equation}
where $\Delta r_i$ is the radial step of cell $i$. Inserting equation
\eqref{eq:cst_aspect_ratio} yields :
\begin{equation}
\label{eq:step_by_step}
r_{i+1}=q r_i \quad \text{where} \quad q{=}\frac{1+\zeta\Delta\theta/2}{1-\zeta\Delta\theta/2} \,.
\end{equation}
Hereafter, $q$ is called the scaling factor. We fix the radii of the inner and outer 
radial boundaries of the domain, $r_{in}$ and $r_{out}$ respectively, along with the total number 
of cells in the radial direction $N$ (excluding the ghost cells). The scaling factor is then computed 
from equation \eqref{eq:step_by_step} applied to the two boundaries: 
$q=\left(r_{out}/r_{in} \right)^{1/N}$.
We can derive the required aspect ratio for a radially stretched grid :
\begin{equation}
\zeta=\frac{2(q-1)}{\Delta\theta(q+1)}.
\end{equation} 
For a stretching smooth enough and a usual number (typically 2) of ghost cells, 
the radial positions of the inner ghost cells remain positive.

As should be clear from our example applications, the stretched grid is created accordingly to the type of symmetry (cylindrical or spherical) and to the dimensionality (1D, 2D or 3D). The numerical schemes 
incorporate the variable radial step size where needed. The coupling between stretched 
grids and block-based AMR has been made possible by computing a scaling factor for each AMR 
level. Since children blocks at the next level span the same radial range as the mother block 
but with twice as many cells, the scaling factor $q_{l+1}$ of the children blocks at level $l+1$ can be 
obtained from the scaling factor $q_{l}$ of the mother block at level $l$ via $q_{l+1}=\sqrt{q_{l}}$. 

The time gain can be estimated by writing the ratio of the number of radial cells in a non-stretched versus a stretched grid, with the same absolute radial step at the inner boundary. It approximately amounts to :
\begin{equation}
\Lambda\sim\frac{r_{out}/r_{in}}{2\log_{10}\left(r_{out}/r_{in}\right)}
\end{equation}
which quickly rises when $r_{out}/r_{in}$ exceeds a few tens. Moreover, in combination with AMR, 
the stretched grid enables to resolve off-centered features (see e.g. our 3D application in Section~\ref{sec:3D_sph_AMR}), without lowering the time step, if the latter is dominated by dynamics 
within the innermost regions.  In the previous section on 1D Bondi accretion, the number of cells for an accurate solution could be reduced by two orders of magnitude.

\subsection{The 2.5D Bondi--Hoyle--Lyttleton problem}
\label{sec:BHL_25D}
A truly two-dimensional accretion-type problem is the Bondi--Hoyle--Lyttleton (BHL) model designed 
by \citet{Hoyle:1939fl} and \citet{Bondi1944} \citep[for a more detailed review of the BHL model and its 
refinements, see][]{Edgar:2004ip,ElMellah2016}. The major change with respect to the 1D Bondi flow is that there is now a supersonic bulk planar motion of the flow with respect to the accretor, at speed $v_{\infty}>c_{\infty}$, which has two serious consequences:
\begin{enumerate}
\item the geometry of the problem is no longer isotropic but is still axisymmetric (around the 
axis through the accretor and with direction set by the speed of the inflow at infinity),
\item since the flow is supersonic at infinity, there can be a bow shock induced by the interaction with the accretor. The isentropic 
assumption we used to bypass the energy equation in section\,\ref{sec:ex_1} no longer holds. 
\end{enumerate}
Analytic considerations point to a formation of the shock at a spatial scale of the 
order of the accretion radius $R_{acc}=2GM_*/v_{\infty}^2$. This radius is close to the 
size of the accretor, when we study stars accreting the wind of their 
companion in binary systems \citep{Chen2017}. However, for compact accretors which have an associated Schwarzschild radius $R_{sch}=2GM_*/c^2$, the contrast between accretor size and the scale at which the flow is deflected by the gravitational influence of the accretor becomes important, since
\begin{equation}
\frac{R_{acc}}{R_{sch}}=\left(\frac{c}{v_{\infty}}\right)^2 \,, \label{accchallenge}
\end{equation}
where $c$ is the speed of light. For realistic speeds $v_\infty$ of stellar winds in wind-dominated X-ray binaries 
for instance, it means that the accretor is 4 to 5 orders of magnitude smaller than the 
shock forming around it. To bridge the gap between those scales, a stretched grid is 
mandatory. 

This has been done with \texttt{MPI-AMRVAC} in~\citet{ElMellah2015}, where we solved 
the hydro equations on a 2D $(r,\theta)$ grid combined with axisymmetry about the ignored $\varphi$ direction. 
The set of equations to solve now include the energy 
equation, also with work done by gravity appearing as a source term, where the energy variable $E=p/(\gamma-1)+\rho(v_r^2+v_\theta^2)/2$ is used to deduce the pressure.  In the simulations, we work with an adiabatic index 
representative of a monoatomic gas, $\gamma=5/3$. The equations solved write as: 
\begin{eqnarray}
\frac{\partial \rho}{\partial t}+\frac{1}{r^2}\frac{\partial (r^2 \rho v_r)}{\partial r} +\frac{1}{r\sin\theta}\frac{\partial(\rho v_\theta \sin\theta)}{\partial \theta} & = & 0 \,, \\
\frac{\partial (\rho v_r)}{\partial t}+\frac{1}{r^2}\frac{\partial (r^2 \rho v_r^2 )}{\partial r} +\frac{1}{r\sin\theta}\frac{\partial(\rho v_r v_\theta\sin\theta)}{\partial\theta} +\frac{\partial p}{\partial r}& = &  \frac{\rho v_\theta^2}{r}- \frac{\rho}{2 r^2} \,, \\
\frac{\partial (\rho v_\theta)}{\partial t}+\frac{1}{r^2}\frac{\partial (r^2 \rho v_r v_\theta )}{\partial r} +\frac{1}{r\sin\theta}\frac{\partial(\rho v_\theta^2 \sin\theta)}{\partial\theta} +\frac{1}{r}\frac{\partial p}{\partial \theta}& = &  -\frac{\rho v_r v_\theta}{r} \,, \\
\frac{\partial E}{\partial t}+\frac{1}{r^2}\frac{\partial (r^2 (E+p) v_r)}{\partial r}+\frac{1}{r\sin\theta}\frac{\partial [(E+p)v_\theta \sin\theta]}{\partial \theta} & = & - \frac{\rho v_r}{2 r^2} \,.
\end{eqnarray}
The terms on the RHS are handled in geometric and gravitational source terms.

Fortunately, not all quantities require an inner $r_{in}$ boundary which matches the 
physical size of the accretor. For instance, \citet{Ruffert1994a} quantified the 
impact of the size of the inner boundary $r_{in}$ on the measured mass accretion rates 
at the inner boundary, $\dot{M}_{in}$ \citep[see Figure 5.3 in][]{ElMellah2016}. The 
result is that $\dot{M}_{in}$ matches the analytic value provided one reaches the 
sonic radius. But as explained in the previous 1D Bondi test, and as can be seen in 
Figure~\ref{fig:rs_gm}, the sonic radius drops to zero for $\gamma=5/3$. \citet{Ruffert1994a} showed that to retrieve the analytic value 
of the mass accretion rate within a few percent, $r_{in}$ must be below a hundredth of 
the accretion radius. Therefore, in the simulations where $\gamma=5/3$, we need to span 
at least 3 orders of magnitude: while the outer boundary lies at $8R_{acc}$, the inner 
boundary has a radius of a thousand of the accretion radius. With a resolution of $176\times64$ cells in the $(r,\theta)$ plane (without the ghost cells), it gives an aspect ratio of $1.04$ (and a scaling factor of $1.05$). 
\begin{figure}
\includegraphics[width=\textwidth]{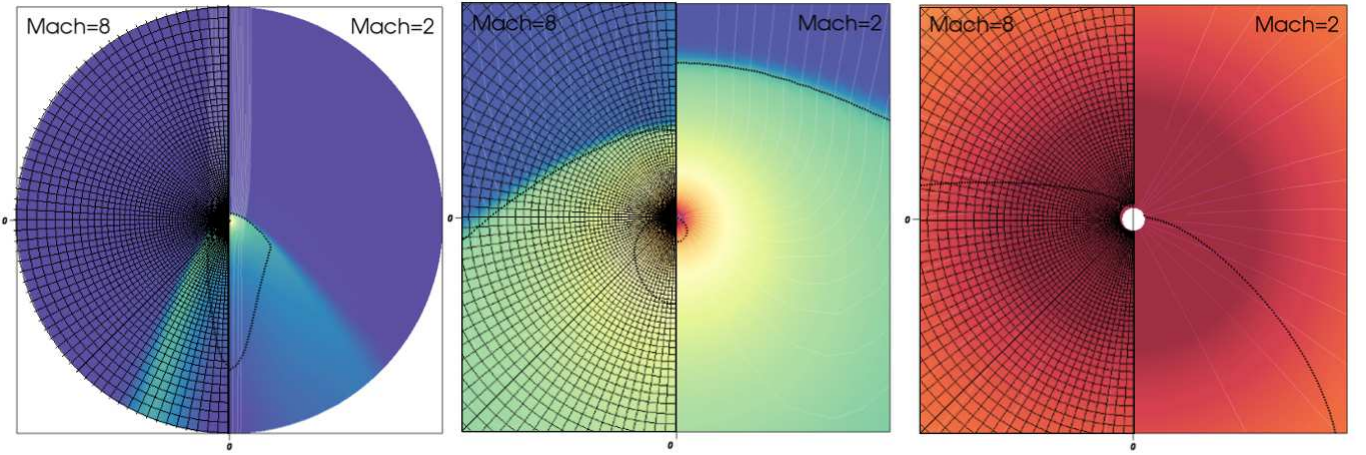}
\caption{Logarithmic colour maps of the density for Mach numbers at infinity of 8 (left 
panels) and 2 (right panels). Some streamlines have been represented (solid white) and 
the two dotted black contours stand for the Mach-1 surface. Zoom in on the central area 
by a factor of 20 (center) and 400 (right).}
\label{fig:zoom}
\end{figure}

The steady solution has the supersonic planar outflow upstream, suddenly transitioning to a subsonic inflow at the bow shock. Then, part of this now subsonic inflow is still accreted in a continuous Bondi-type transonic fashion in the wake of the accretor. Solutions for different Mach numbers $\mathcal{M}_{\infty}=v_\infty/c_\infty$ of the supersonic planar inflow are compared in~Figure~\ref{fig:zoom}, showing density variations in a zoomed in fashion. 
The upstream conditions at $8R_{acc}$ are set by the ballistic solutions 
\citep{Bisnovatyi-Kogan1979} and the downstream boundary condition is continuous (with a supersonic 
outflow). At the inner boundary, we enforce the same boundary conditions as described in section\,\ref{sec:ex_1}, with a continuous $v_{\theta}$. \texttt{MPI-AMRVAC} provides a steady solution on a stretched (non AMR) grid which 
matches the orders of magnitude mentioned above: the position of the detached bow shock, 
the jump conditions at the shock, as well as the inner mass accretion rate \citep{ElMellah2015}. 
In particular, we retrieve a topological property derived by \citet{Foglizzo1996} 
concerning the sonic surface, deeply embedded in the flow and anchored into the inner 
boundary, whatever its size (see the solid black contour denoting its location in the 
zoomed views in Figure~\ref{fig:zoom}). 

\subsection{3D Clumpy wind accretion in Supergiant X-ray Binaries}
\label{sec:3D_sph_AMR}

A third hydrodynamic application of radially stretched AMR grids generalizes the 
2D (spherical axisymmetric) setup above to a full 3D case. In a forthcoming paper
(El Mellah et al., in prep.), we use a full 3D spherical mesh to study wind-accretion 
in Supergiant X-ray Binaries, where a compact object orbits an evolved mass-losing star. In these 
systems, the same challenge of scales identified by equation~(\ref{accchallenge}) 
is at play, and the BHL shock-dominated, transonic flow configuration is applicable. 
In realistic conditions, the wind of the evolved donor star is radiatively 
driven and clumpy, making the problem fully time-dependent. We model the accretion 
process of these clumps onto the compact object, where the inhomogeneities (clumps) 
in the stellar wind  enter a spherically stretched mesh centered on the accreting 
compact object, and the clump impact on the time-variability of the mass accretion 
rate is studied. The physics details will be provided in that paper, here we 
emphasize how the 3D stretched AMR mesh is tailored to the problem at hand. The 
initial and boundary conditions are actually derived from the 2.5D solutions 
described above, while a seperate 2D simulation of the clump formation and 
propagation in the companion stellar wind serves to inject 
time-dependent clump features in the supersonic inflow.

\begin{figure}
\includegraphics[width=\textwidth]{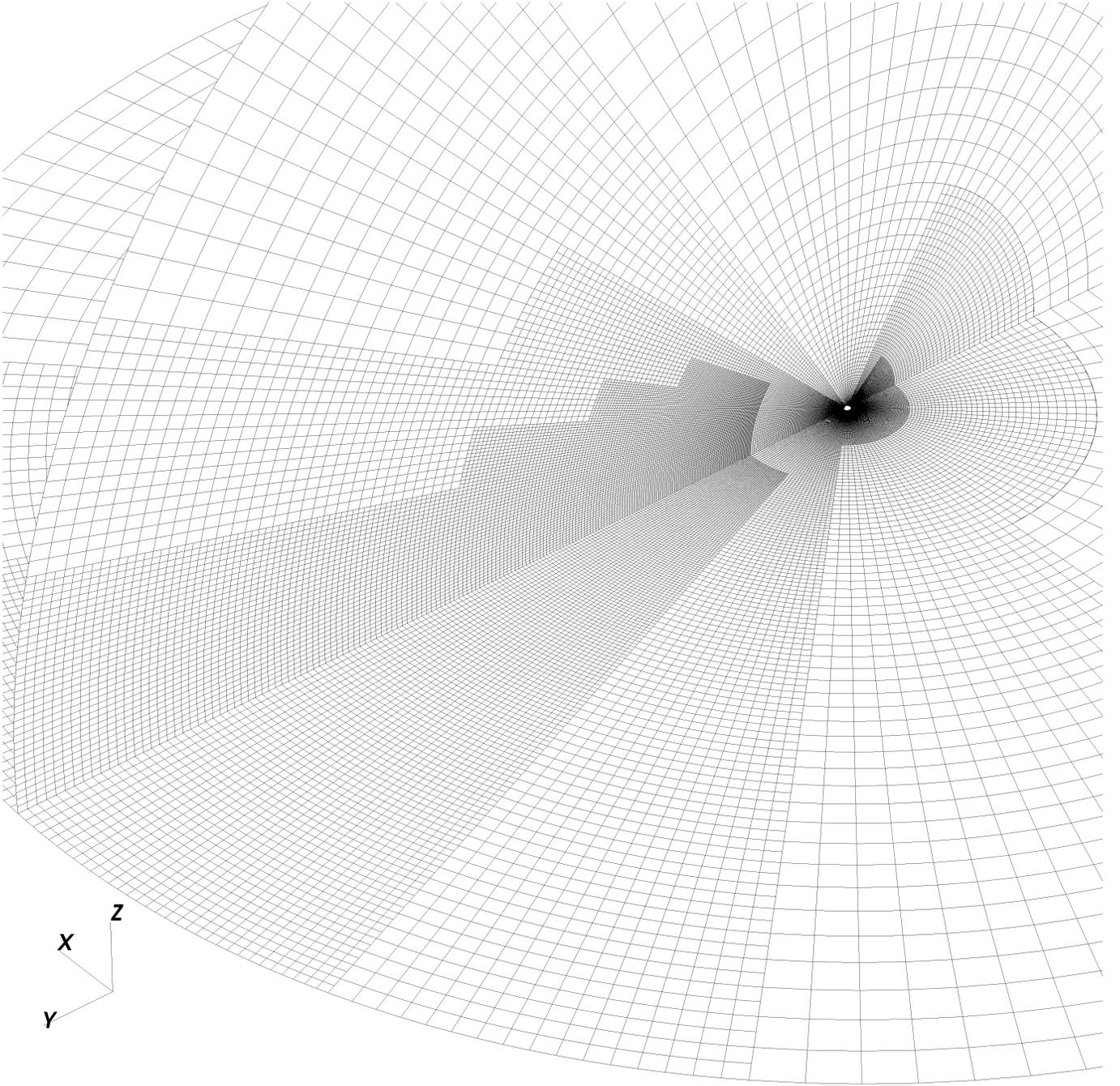}	
\caption{Two slices of the upper hemisphere of a spherical mesh with 4 levels of 
refinement, as used in a study of time-dependent wind-accretion onto a compact object. 
User-defined restrictions on AMR prevent the grid to be refined in some 
areas (e.g. in the downstream hemisphere, where the $y$-coordinate is negative) far from the inner 
boundary and in the vicinity of the polar $z$-axis). We further restrict the maximum level 
of refinement to 3 in the vicinity of the inner boundary and allow for maximum refinement in 
the outer regions of the upstream hemisphere close to the $y$-axis. This corresponds to 
the supersonic inflow area where clumps enter the grid.}
\label{fig:mesh_str_AMR}
\end{figure} 

The use of a 3D spherical mesh combined with explicit time stepping brings in 
additional challenges since the time step $\Delta t$ must respect the Courant--Friedrich--Lewy 
(CFL) condition. Since the azimuthal line element is $r\sin\theta\Delta\varphi$ with colatitude $\theta$, it means that the time step obeys:
\begin{equation}
\Delta t \left(\theta\right)\propto \sin\left(\theta\right) \,.
\end{equation}
The axisymmetry adopted in the previous section avoided this complication.  As a consequence, the additional cost in terms of number of iterations to run 
a 3D version of the setup presented in the previous section is approximately:
\begin{equation}
\frac{\Delta t \left(\pi/2\right)}{\Delta t \left(\theta_{min}\right)} \sim \frac{1}{\theta_{min}} \,,
\end{equation}
where $\theta_{min}\ll \pi/2$ is the colatitude of the first cell center just off the polar axis. For a 
resolution on the base AMR level of 64 cells spanning $\theta=0$ to $2\pi$, 
it means a time step $40$ times smaller compared to the two-dimensional counterpart. Together with the 
additional cells in the azimuthal direction, this makes a 3D setup much more computationally demanding. 

We compromise on this aspect by using an AMR mesh which deliberately exploits  a lower 
resolution at the poles, while still resolving all relevant wind structures. Since the clumps embedded in the supersonic inflow are small scale structures, we need to enable AMR (up to 4 levels in the 
example of Figure~\ref{fig:mesh_str_AMR}) in the upstream hemisphere to resolve them, 
in particular at the outer edge of the mesh where the radially stretched cells have 
the largest absolute size on the first AMR level. At the same time, we are only 
interested in the fraction of the flow susceptible to be accreted by the compact object, so we can inhibit AMR refinement in the downstream hemisphere, except in the immediate 
vicinity of the accretor (below the stagnation point in the wake). We also prevent 
excessive refinement in the vicinity of the accretor (no refinement beyond the third 
level in Figure~\ref{fig:mesh_str_AMR}) since the stretching already provides more 
refined cells. Finally, in the upstream hemisphere, we favor AMR refinement in the 
accretion cylinder, around the axis determined by the direction of the inflow velocity. Note that this axis was the symmetry polar axis in the previous 2.5D setup in section \ref{sec:BHL_25D}, but this is different in the present 3D setup, where the polar axis is above and below the central accretor. In the application to study supergiant X-ray binary accretion, together with geometrically enforced AMR, we use dynamic AMR with a refinement criterion on the mass density to follow the flow as it is accreted onto the compact object.

\subsection{Trans-Alfv\'enic solar wind}

\begin{figure}
\includegraphics[width=\textwidth]{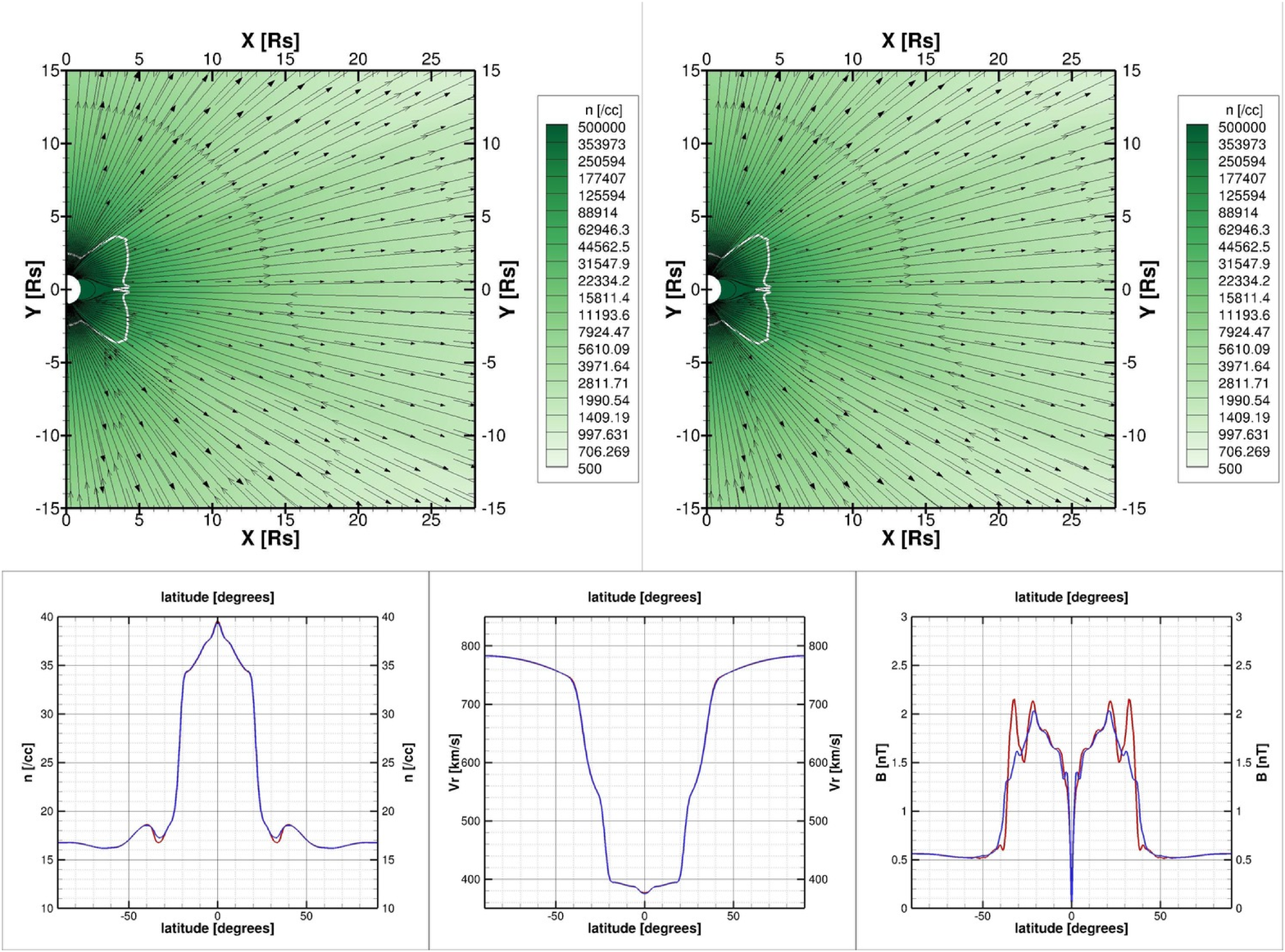}
\caption{Axisymmetric 2.5D MHD simulations of the solar wind. The top panels show the color-coded density, the velocity vectors, the magnetic 
field lines (in black), as well as the Alfv\'en surface (white line) for two simulations: one where the magnetic field splitting method 
is used (right panel) and one where it is not (left panel).
The lower panels show the solar wind characteristics at 1~AU: the plasma number density (left panel), radial velocity (middle panel),
and the magnetic field magnitude (right panel) for both simulation (with MFS in red and without MFS in blue).}
\label{fig:SW}
\end{figure}

To simulate the propagation of the solar wind between the Sun and the orbit of the Earth (at 215~R$_\odot$), the use of a spherical stretched grid 
can also be very important. First, it considerably reduces the amount of cells needed in the simulations (by a factor of 46). 
Second, it allows to maintain the same aspect ratio for all the cells in the simulation. We present here the results of 2.5D MHD 
simulations of the meridional plane of the solar wind, i.e., axisymmetric simulations on a 2D mesh where the 3 vector components are considered
for the plasma velocity and for the magnetic field. This simulation setup also allows us to test the magnetic field splitting (MFS) method discussed in 
Section~\ref{sec:b0field}. Instead of solving the induction equation for the total magnetic field ($\mathbf{B}$), one can solve it for the difference with 
the internal dipole field of the Sun ($\mathbf{B_1}=\mathbf{B}-\mathbf{B_0}$), where $\mathbf{B_0}$ represent the dipole field. We will here compare 
the results of these two setups.
In our simulations, the solar wind is generated through the inner boundary (at $r$=1~R$_\odot$), in a fashion similar to 
\citet{Jacobs2005,Chane2005,Chane2008}: the temperature is fixed to $1.5\times10^6$ K, and the number density to $7.1\times10^6$ cm$^{-3}$, 
where the azimuthal velocity is set to 
$(2.09-0.264 \cos^2\theta - 0.334 \cos^4\theta)\sin\theta$ km s$^{-1}$ to enforce the solar differential 
rotation of the Sun ($\theta$ is the co-latitude), and where the radial component of the magnetic field is fixed to the 
internal dipole value (with $B_r=1.1$ G at the equator).
In addition, close to the equator (at latitudes lower than $22.6^\circ$), in the streamer region, a dead-zone is enforced at the boundary
by fixing the radial and the latitudinal velocities to zero. 
The equations solved are the ideal MHD equations plus two source terms: one for the heating of the corona and one for the gravity of the Sun.
The heating of the corona is treated as an empirical source term in the energy equation. We follow the procedure described in \citet{Groth2000}
with a heating term dependent on the latitude (with a stronger heating at high latitudes)
and on the radial distance (the heating decreases exponentially). The inclusion of this source term result in a slow solar wind close to the
equatorial plane and a fast solar wind at higher latitudes (see Figure~\ref{fig:SW})

The simulations are performed on a spherical stretch mesh between 1~R$_\odot$ and 216~R$_\odot$ with 380 cells in the radial direction and
224 cells in the latitudinal direction. For these simulations, we do not take advantage of the AMR capabilities of the code and one single mesh level
is used (no refinement). The smallest cells (close to the sun) are 0.014~R$_\odot$ large, while the largest cells (at 1~AU) are 3~R$_\odot$ large.
The equations are solved with a finite-volume scheme, using the two-step HLL solver \citep{Harten83} with Koren slope limiters.
Powell's method \citep{Powell1999} is used for the divergence cleaning of the magnetic field.
The adiabatic index $\gamma$ is set to $5/3$ in these simulations.

The results of the simulations are shown in Figure~\ref{fig:SW}. The top panels display the plasma density, the velocity vectors,
the magnetic field lines, and the Alfv\'en surface. One can see that both panels are very similar with: 
1) a fast solar wind at high latitudes and a slow denser wind close to the equatorial plane, 
2) a dense helmet streamer at low latitudes close to the Sun where the plasma velocity is almost zero,
3) an Alfv\'en surface located between 2 and 7~R$_\odot$ in both cases.
To highlight the differences between the two simulations, one dimensional cuts at 1~AU are provided in the lower panels of 
Figure~\ref{fig:SW} for both simulations. 
The density profiles are very similar, with only small differences at the transition between the slow and the fast solar winds.
For both simulations, the solar wind is, as expected, denser close to the equatorial plane. 
The radial velocity profiles (shown in the middle panel) are nearly identical in both simulations, with a fast solar wind 
(a bit less than 800~km s$^{-1}$) at high latitudes, and a slow solar wind (a bit less than 400~km s$^{-1}$) close to the equator.
These results are very similar to in situ measurements at 1~AU \citep[see, for instance,][]{Phillips1995}.
The differences are more pronounced for the magnetic field magnitude (right panel), where the magnetic field can be up to 30\% stronger
in a small region located at the transition region between the fast and slow solar winds when the MFS method is used.
In general, the differences between the two simulations are negligible and confined to small specific areas of the simulations.

\section{Magnetic field splitting} \label{sec:b0field}

When numerically solving for dynamics in solar or planetary atmospheres, the strong magnetic 
field and its gradients often cause serious difficulties if the total magnetic field 
is solved for as a dependent variable. This is especially true when both high- and low-$\beta$ 
regions need to be computed accurately in shock-dominated interactions. An elegant 
solution to this problem~\citep{Tanaka94} is to split the magnetic field 
$\mathbf{B}=\mathbf{B_0}+\mathbf{B_1}$, into a dominating background field $\mathbf{B_0}$ 
and a deviation $\mathbf{B_1}$, and solve a set of modified equations directly 
quantifying the evolution of $\mathbf{B_1}$. Splitting off this background magnetic field component and 
only evolving the perturbed magnetic field is more accurate and robust, in the sense 
that it diminishes the artificial triggering of negative gas pressure in extremely low-$\beta$ plasma. This background field 
$\mathbf{B_0}$ was originally limited to be a time-independent potential field \citep{Tanaka94} 
(as used in the previous section for the solar wind simulation on a stretched mesh), and 
in this form has more recently led to modify the formulations of HLLC~\citep{Guo15} or 
HLLD~\citep{Guo16} MHD solvers, which rely on a 4 to 6-state simplification of the local 
Riemann fans, respectively. The splitting can also be used for handling background 
time-dependent, potential fields~\citep{Toth08}, even allowing for extensions to 
Hall-MHD physics where challenges due to the dispersive whistler wave dynamics enter. 
The governing equations for a more general splitting, as revisited below, were also 
presented in~\citet{Gombosi02}, giving extensions to semi-relativistic MHD where the 
Alfv\'en speed can be relativistic. The more general splitting form also appears 
in~\citet{Feng11}, although they further apply it exclusively with time-independent, 
potential fields.

\subsection{Governing equations, up to resistive MHD}
Here, we demonstrate this MFS method in applications that allow any kind of 
time-independent magnetic field as the background component, no longer restricted 
to potential (current-free) settings. At the same time, we wish to also allow for 
resistive MHD applications that are aided by this more general MFS technique. The 
derivation of the governing equations employing an arbitrary split in a 
time-independent $\mathbf{B_0}$ are repeated below,  to extend them further on to 
resistive MHD. The ideal MHD equations in conservative form are given by
\begin{align}
 \frac{\partial \rho}{\partial t}+\nabla\cdot\left(\rho\mathbf{v}\right)&=0, \\
 \frac{\partial \left(\rho\mathbf{v}\right)}{\partial t}+\nabla\cdot\left(
 \rho\mathbf{vv}+(p+\frac{\mathbf{B}\cdot\mathbf{B}}{2\mu_0})\mathbf{I}-\frac{\mathbf{BB}}{\mu_0}\right)&=0,  \label{eq:mom}\\
 \frac{\partial E}{\partial t}+\nabla\cdot\left(E\mathbf{v}+(p+\frac{\mathbf{B}\cdot\mathbf{B}}{2\mu_0})\mathbf{v}-
 \frac{\mathbf{B}}{\mu_0}(\mathbf{B}\cdot\mathbf{v})\right)&=0,  \label{eq:energy}\\
 \frac{\partial \mathbf{B}}{\partial t}+\nabla\cdot\left(\mathbf{vB}-\mathbf{Bv}\right)&=0,  \label{eq:mag}
\end{align}
where $\rho$, $\mathbf{v}$, $\mathbf{B}$, and $\mathbf{I}$ are the plasma
density, velocity, magnetic field, and unit tensor, respectively. In solar applications, 
gas pressure is $p=2.3 n_{\rm  H} k_{\rm B} T$, assuming full ionization and a solar 
hydrogen to helium abundance $n_{\rm He}/n_{\rm H}=0.1$, where $\rho=1.4 n_{\rm H} m_p$ (or the mean molecular weight $\tilde{\mu}\approx 0.6087$),  while
$E=p/(\gamma-1)+\rho v^2/2+B^2/2\mu_0$ is the total energy.
Consider $\mathbf{B}=\mathbf{B_0}+\mathbf{B_1}$, where $\mathbf{B_0}$ is a time-independent 
magnetic field, satisfying 
$\partial \mathbf{B_0}/\partial t=0$ and $\nabla\cdot\mathbf{B_0}=0$.
Equation (\ref{eq:mom}) can be written as
\begin{gather}
 \frac{\partial \left(\rho\mathbf{v}\right)}{\partial t}+\nabla\cdot\left(
 \rho\mathbf{vv}+(p+\frac{B_1^2}{2\mu_0})\mathbf{I}-\frac{\mathbf{B_1B_1}}{\mu_0}\right)+\notag\\
  \nabla\cdot\left(\frac{\mathbf{B_0}\cdot\mathbf{B_1}}{\mu_0}\mathbf{I}-\frac{\mathbf{B_0B_1}+
   \mathbf{B_1B_0}}{\mu_0} \right)=\frac{1}{\mu_0}(\nabla\times\mathbf{B_0})\times\mathbf{B_0}. \label{eq:mom2}
\end{gather}
Defining $E_1\equiv p/(\gamma-1)+\rho v^2/2+B_1^2/(2\mu_0)$, then 
$E=E_1+(B_0^2+2\mathbf{B_0}\cdot\mathbf{B_1})/(2\mu_0)$ and equation~(\ref{eq:energy}) can be written as
\begin{gather}
 \frac{\partial E_1}{\partial t}+\frac{\mathbf{B_0}}{\mu_0}\cdot\frac{\partial\mathbf{B_1}}{\partial t}+
 \nabla\cdot\left(E_1\mathbf{v}+(p+\frac{B_1^2}{2\mu_0})\mathbf{v}-\frac{\mathbf{B_1}}{\mu_0}(\mathbf{B_1}\cdot\mathbf{v})\right)+\notag\\
 \nabla\cdot\left(\frac{\mathbf{B_0}\cdot\mathbf{B_1}}{\mu_0}\mathbf{v}-\frac{\mathbf{B_0}}{\mu_0}(\mathbf{B_1}\cdot\mathbf{v})\right)+
 \nabla\cdot\left(\frac{\mathbf{B_0}\cdot\mathbf{B}}{\mu_0}\mathbf{v}-\frac{\mathbf{B}}{\mu_0}(\mathbf{B_0}\cdot\mathbf{v})\right)=0. \label{eq:e1}
\end{gather}
When we take the scalar product of equation~(\ref{eq:mag}) with a time-independent $\mathbf{B_0}$, we can use vector manipulations on the right-hand side (RHS) of 
$ \mathbf{B_0}\cdot{\partial \mathbf{B_1}}/{\partial t}=\mathbf{B_0}\cdot\nabla\times(\mathbf{v}\times\mathbf{B})$ to obtain
 \begin{gather}
 \frac{\mathbf{B_0}}{\mu_0}\cdot\frac{\partial\mathbf{B_1}}{\partial t}+
 \nabla\cdot\left(\frac{\mathbf{B_0}\cdot\mathbf{B}}{\mu_0}\mathbf{v}-\frac{\mathbf{B}}{\mu_0}(\mathbf{B_0}\cdot\mathbf{v})\right)=
   (\mathbf{v}\times\mathbf{B})\cdot\frac{1}{\mu_0}\nabla\times\mathbf{B_0} \,.\label{eq:B1}
\end{gather}
Inserting equation (\ref{eq:B1}) into equation (\ref{eq:e1}), we arrive at
\begin{gather}
 \frac{\partial E_1}{\partial t}+
 \nabla\cdot\left(E_1\mathbf{v}+(p+\frac{B_1^2}{2\mu_0})\mathbf{v}-\frac{\mathbf{B_1}}{\mu_0}(\mathbf{B_1}\cdot\mathbf{v})\right)+\notag\\
 \nabla\cdot\left(\frac{\mathbf{B_0}\cdot\mathbf{B_1}}{\mu_0}\mathbf{v}-\frac{\mathbf{B_0}}{\mu_0}(\mathbf{B_1}\cdot\mathbf{v})\right)
 =-(\mathbf{v}\times\mathbf{B})\cdot\frac{1}{\mu_0}\nabla\times\mathbf{B_0}.\label{eq:e1new}
\end{gather}
Therefore, in the ideal MHD system with MFS, we see how a source term appears in the RHS for the momentum equation~(\ref{eq:mom2}) that quantifies (partly) the Lorentz force $\mathbf{J_0}\times\mathbf{B_0}$,  where $\mathbf{J_0}=\nabla\times\mathbf{B_0}/{\mu_0}$, while the energy equation~(\ref{eq:e1new}) for $E_1$ features a RHS written as $\mathbf{E}\cdot\mathbf{J_0}$ for the electric field $\mathbf{E}=-\mathbf{v}\times\mathbf{B}$. The induction equation, for a time-invariant $\mathbf{B_0}$ just writes as before:
\begin{gather}
 \frac{\partial \mathbf{B_1}}{\partial t}+\nabla\cdot(\mathbf{vB_1}-\mathbf{B_1v})+\nabla\cdot(\mathbf{vB_0}-\mathbf{B_0v})=\mathbf{0} \,. \label{eq:Bs}
\end{gather}
For force-free magnetic fields, $\nabla\times\mathbf{B_0}=\alpha\mathbf{B_0}$, 
where $\alpha$ is constant along one magnetic field line, we have $\mathbf{J_0}\times\mathbf{B_0}=\mathbf{0}$ 
and $-(\mathbf{v}\times\mathbf{B})\cdot\mathbf{J_0}=-(\mathbf{v}\times\mathbf{B_1})\cdot\mathbf{J_0}$.

For resistive MHD, extra source terms $\nabla\cdot(\mathbf{B}\times\eta\mathbf{J})/{\mu_0}$ and 
$-\nabla\times(\eta\mathbf{J})$, are added on the RHS of the energy equation~(\ref{eq:energy}) 
and the induction equation~(\ref{eq:mag}), respectively, where $\mathbf{J}=\nabla\times\mathbf{B}/{\mu_0}$.
Then the energy equation must account for Ohmic heating and magnetic field diffusion, to be written as
\begin{gather}
 \frac{\partial E_1}{\partial t}+\frac{\mathbf{B_0}}{\mu_0}\cdot\frac{\partial\mathbf{B_1}}{\partial t}+
 \nabla\cdot\left(E_1\mathbf{v}+(p+\frac{B_1^2}{2\mu_0})\mathbf{v}-\frac{\mathbf{B_1}}{\mu_0}(\mathbf{B_1}\cdot\mathbf{v})\right)+ \notag\\
 \nabla\cdot\left(\frac{\mathbf{B_0}\cdot\mathbf{B_1}}{\mu_0}\mathbf{v}-\frac{\mathbf{B_0}}{\mu_0}(\mathbf{B_1}\cdot\mathbf{v})\right)+
 \nabla\cdot\left(\frac{\mathbf{B_0}\cdot\mathbf{B}}{\mu_0}\mathbf{v}-\frac{\mathbf{B}}{\mu_0}(\mathbf{B_0}\cdot\mathbf{v})\right) \notag\\
 =\eta J^2-\frac{1}{\mu_0}\mathbf{B_0}\cdot\nabla\times(\eta\mathbf{J}) -\frac{1}{\mu_0}\mathbf{B_1}\cdot\nabla\times(\eta\mathbf{J}).  \label{eq:e1r}
\end{gather}
The induction equation, extended with resistivity, now gives
\begin{gather}
 \frac{\mathbf{B_0}}{\mu_0}\cdot\frac{\partial\mathbf{B_1}}{\partial t}+
 \nabla\cdot\left(\frac{\mathbf{B_0}\cdot\mathbf{B}}{\mu_0}\mathbf{v}-\frac{\mathbf{B}}{\mu_0}(\mathbf{B_0}\cdot\mathbf{v})\right)= \notag\\
   (\mathbf{v}\times\mathbf{B})\cdot\frac{1}{\mu_0}\nabla\times\mathbf{B_0} -\frac{1}{\mu_0}\mathbf{B_0}\cdot\nabla\times(\eta\mathbf{J}). \label{eq:B1r}
\end{gather}
Insert equation (\ref{eq:B1r}) to equation (\ref{eq:e1r}), and we have the resistive MHD energy equation in MFS representation
\begin{gather}
 \frac{\partial E_1}{\partial t}+
 \nabla\cdot\left(E_1\mathbf{v}+(p+\frac{B_1^2}{2\mu_0})\mathbf{v}-\frac{\mathbf{B_1}}{\mu_0}(\mathbf{B_1}\cdot\mathbf{v})\right)+ \notag\\
 \nabla\cdot\left(\frac{\mathbf{B_0}\cdot\mathbf{B_1}}{\mu_0}\mathbf{v}-\frac{\mathbf{B_0}}{\mu_0}(\mathbf{B_1}\cdot\mathbf{v})\right) \notag\\
  =-(\mathbf{v}\times\mathbf{B})\cdot\mathbf{J_0} +\eta J^2-\frac{1}{\mu_0}\mathbf{B_1}\cdot\nabla\times(\eta\mathbf{J}).
\end{gather}
The induction equation as implemented, is then written as
\begin{gather}
 \frac{\partial \mathbf{B_1}}{\partial t}+\nabla\cdot(\mathbf{vB_1}-\mathbf{B_1v})+\nabla\cdot(\mathbf{vB_0}-\mathbf{B_0v})=
  -\nabla\times(\eta\mathbf{J}). \label{eq20}
\end{gather}
In the implementation of the MFS form of the MHD equations in \texttt{MPI-AMRVAC}, we 
take the following strategies. 
Each time when creating a new block, a special user subroutine is called to calculate $\mathbf{B_0}$ 
at cell centers and cell interfaces which is stored in an allocated array,
and then $\mathbf{J_0}$ is evaluated and stored at cell centers either 
numerically or analytically with an optional user subroutine. 
Note that when the block is deleted from the AMR mesh, these stored values are also lost.
The third term in the left hand side of the momentum equation~(\ref{eq:mom2}), energy equation~(\ref{eq:e1new}) or~(\ref{eq:e1r}) and in the 
induction equation~(\ref{eq:Bs}) or~(\ref{eq20}) is 
added in the flux computation for the corresponding variable, based on pre-determined $\mathbf{B_0}$ 
magnetic field values at the cell interfaces where fluxes are evaluated. The terms in the right 
hand sides of all equations are treated as unsplit source terms, which are added in each sub-step of a
time integration scheme. We use simple central differencing
to evaluate spatial derivatives at cell centers. For example, to compute 
$\nabla\times(\eta\mathbf{J})$ in the physical region of a block excluding ghost cells, we first 
compute $\mathbf{J_1}$ in a block region including the first layer of ghost cells, which needs $\mathbf{B_1}$ at the second layer of ghost cells.

\subsection{MFS applications and test problems}\label{sec:MFStest}

We present three tests and one application to demonstrate the validity and advantage of 
the MFS technique. All three tests are solved with a finite-volume scheme setup combining the 
HLL solver \citep{Harten83} with \v{C}ada's compact third-order 
limiter \citep{Cada09} for reconstruction and a three-step Runge--Kutta time integration in 
a dimensionally unsplit approach. We use a diffusive term in the induction equation to keep the
divergence of magnetic field under control \citep{Keppens03,Holst07,Xia14}. We adopt a constant 
index of $\gamma=5/3$. Unless stated explicitly, we employ dimensionless (code) units everywhere, 
also making $\mu_0=1$. 

\subsubsection{3D force-free magnetic flux rope}
\begin{figure}
\includegraphics[width=\textwidth]{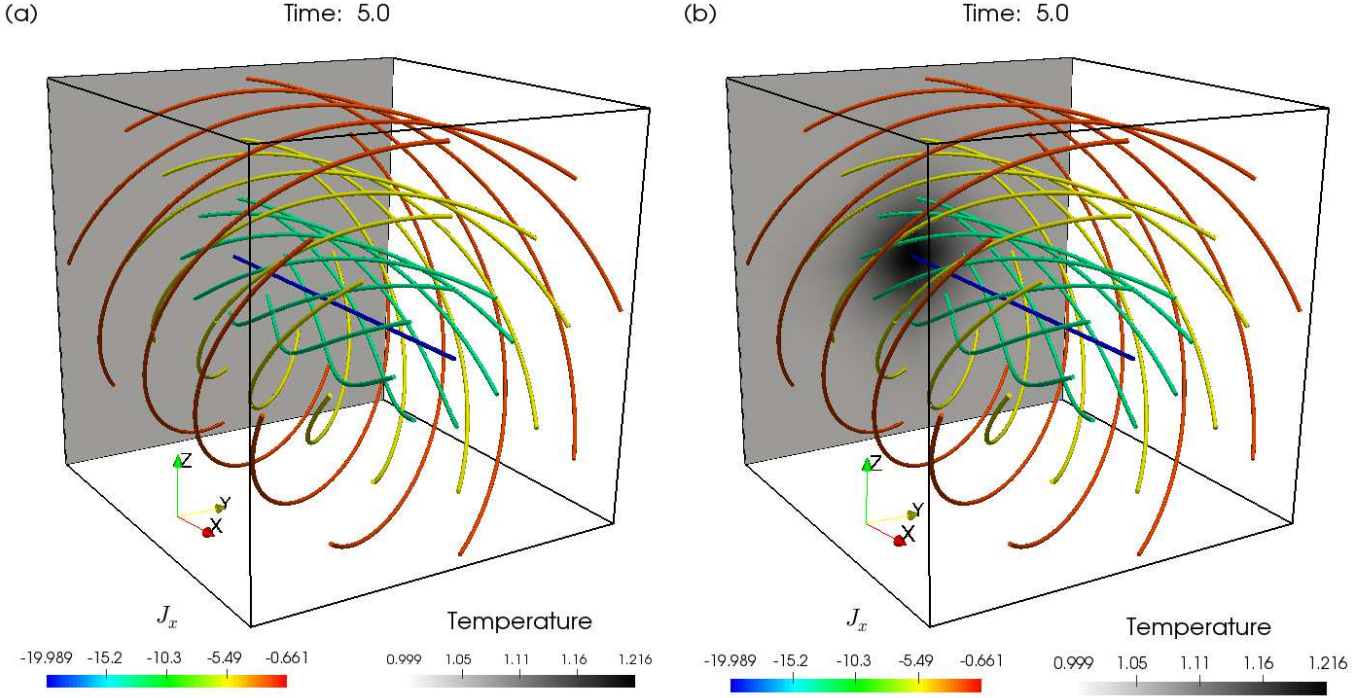}
\caption{3D views of the static magnetic flux rope equilibrium after time 5 with (a) and without (b) 
magnetic field split (MFS), where magnetic field lines are colored by the $x$-component of their
current density and the back cross-sections are colored by temperature.}
\label{fig:FR3d}
\end{figure}

This is a test initialized with a static plasma threaded by
a force-free, straight helical magnetic flux rope on a Cartesian grid (see Figure~\ref{fig:FR3d}). 
The initial density and gas pressure are uniformly set to unity. The simulation box has a size of
 6 by 6 by 6 and is resolved by a $128\times128\times128$ uniform mesh which is decomposed into 
blocks of $16\times16\times16$ cells.
The magnetic field is non-potential and nonlinearly force-free, as proposed by 
Low~\citep{Low77}, and is formulated as
\begin{gather}
B_x=\frac{4\xi}{1+\xi^2}f(y,z),
B_y=2\left(kz+\frac{1-\xi^2}{1+\xi^2}\right)f(y,z),
B_z=-2ky f(y,z),\\
f(y,z)=B_a\left[\frac{4\xi^2}{(1+\xi^2)^2}+k^2y^2+(kz+\frac{1-\xi^2}{1+\xi^2})^2\right]^{-1}.
\end{gather}
We choose free parameters $\xi=1$ and $k=0.5$ to make the axis of the flux rope
align with the $x$-axis and set $B_a=10$. 
The electric current of this flux rope is analytically found to be
\begin{gather}
J_x=4kf(y,z)+4kB_a^{-1}f(y,z)^2\left[k^2y^2+(kz+\frac{1-\xi^2}{1+\xi^2})^2\right],\notag\\
J_y=\frac{8\xi k}{1+\xi^2}B_a^{-1}f(y,z)^2\left(kz+\frac{1-\xi^2}{1+\xi^2}\right),\notag\\
J_z=-\frac{8\xi k}{1+\xi^2}B_a^{-1}f(y,z)^2y.
\end{gather}
The plasma $\beta$ ranges from 0.006 to 0.027. The Alfv\'{e}n crossing time along $y$-axis across 
the box is 0.377, determined by the averaged Alfv\'{e}n speed along $y$-axis. 
This analytical force-free static equilibrium should remain static for ever. However, due to 
the finite numerical diffusion of the numerical scheme adopted, the current in the flux rope slowly 
dissipates when we do not employ the MFS strategy, thereby converting magnetic energy to internal 
energy and resulting in (small but) non-zero velocities. After 5 (dimensionless code) time units (about
8500 CFL limited timesteps), as 
shown in Figure~\ref{fig:FR3d}, the reference run without MFS has up to 20\% increase in temperature 
near the central axis of the flux rope, while the run with 
MFS perfectly maintains the initial temperature. Both runs do not have discernible change in magnetic 
structure.

\begin{figure}
\includegraphics[width=\textwidth]{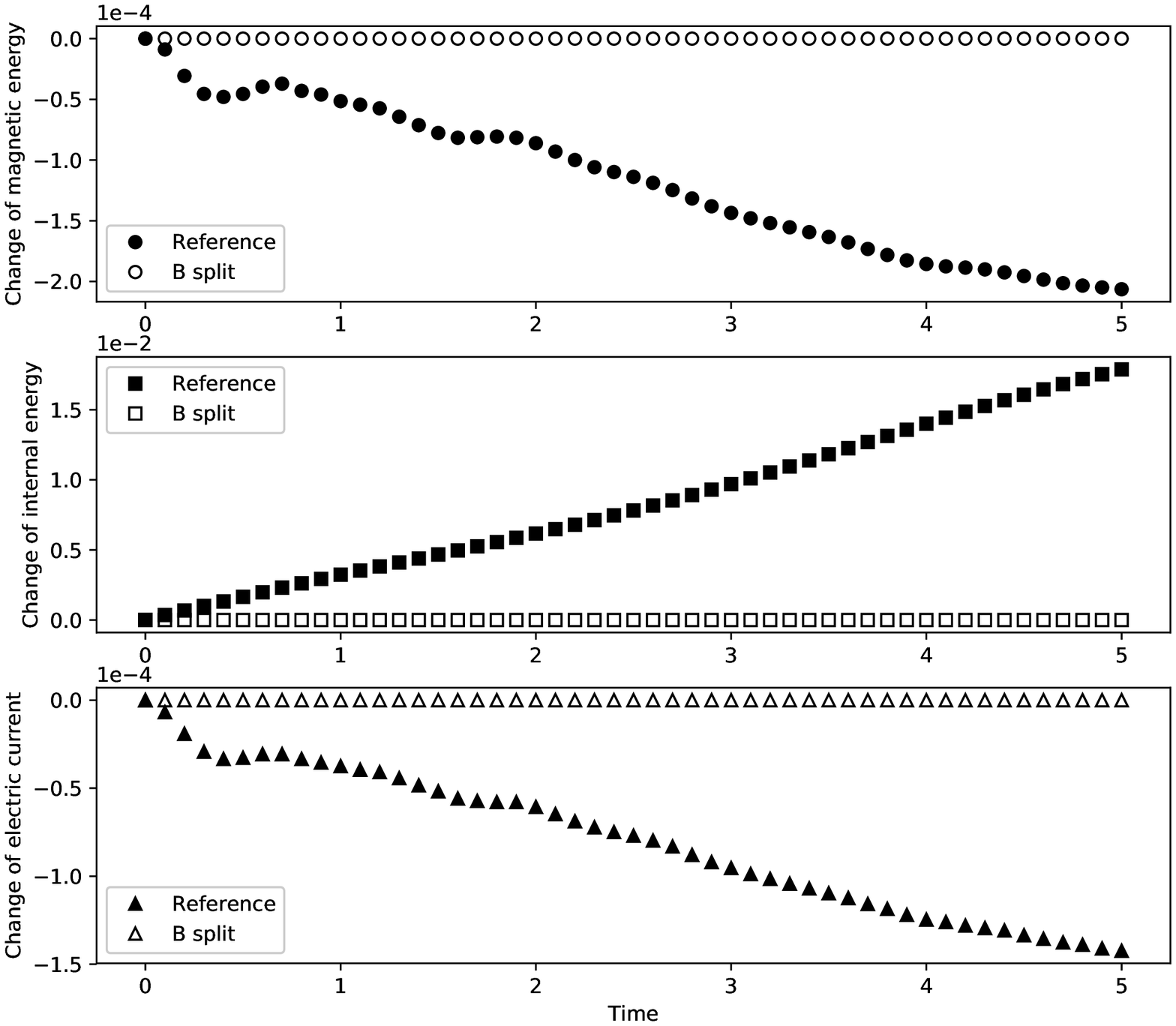}
\caption{Relative changes of total magnetic energy (top), electric current (middle), 
and internal energy (bottom) in the run with MFS (B split) shown by hollow symbols and in 
the reference run shown by filled symbols.}
\label{fig:FRchange}
\end{figure}
We quantify the relative changes, $(f(t)-f(0))/f(0)$, of total magnetic energy, 
electric current, and internal energy in time for both runs and plot them in 
Figure~\ref{fig:FRchange}. In the reference run without MFS, the magnetic energy decreases
up to 0.02 \%, current decreases 0.15 \%, and internal energy increases 1.8 \% at 5 time units. 
Thus, the dissipation of magnetic energy and current and the resulting increase of internal 
energy with the employed discretization is significant for long-term runs. However, the run with the MFS 
strategy keeps the initial equilibrium perfectly and all relative changes are exactly zero. 

\subsubsection{2D non-potential magnetic null-point}
\begin{figure}
\includegraphics[width=\textwidth]{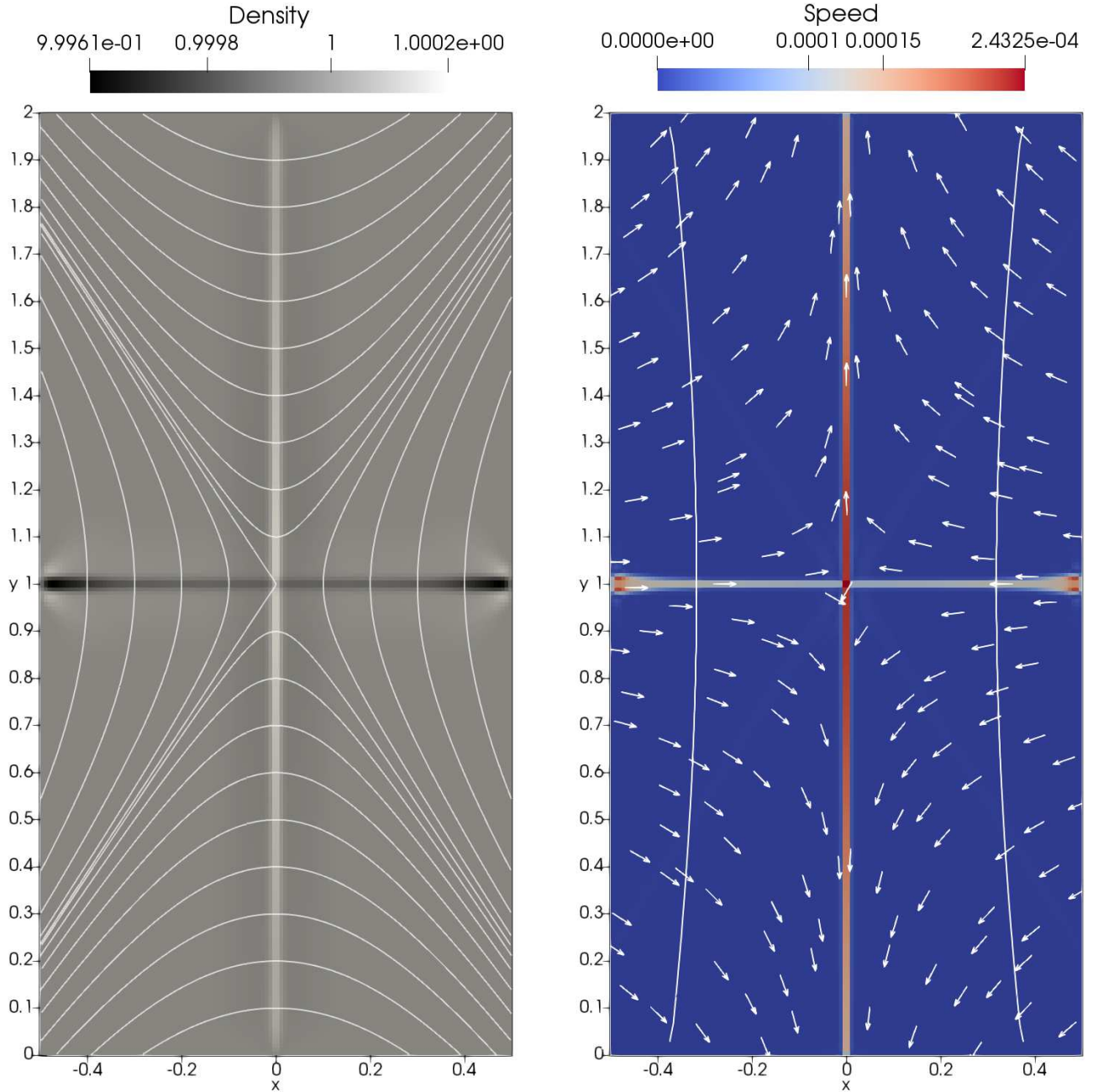}
\caption{Density map with over-plotted magnetic field lines (left) and speed map 
with arrows indicating the local directions of Lorentz force (right) and solid lines where 
plasma $\beta=1$, at 10 time units with MFS.}
\label{fig:nullimage}
\end{figure}

This second test focuses on using the MFS technique in a setup where a magnetic null point 
is present (where magnetic field locally vanishes) in a 2D planar configuration. Using a 
magnetic vector potential $\mathbf{A}=[0,0,A_z]$ with
$A_z=0.25 B_c [(j_t-j_z) (y-y_c)^2-(j_t+j_z) (x-x_c)^2]$, the resulting magnetic field works out to
\begin{gather}
 B_x=0.5 B_c (j_t-j_z) (y-y_c) \,,\notag\\
 B_y=0.5 B_c (j_t+j_z) (x-x_c) \,,\notag\\
 B_z=0\,.
\end{gather}
Its current density is
\begin{gather}
 J_x=J_y=0\,,\notag\\
 J_z=B_c j_z \,.
\end{gather} 
This magnetic field has a 2D non-potential null-point in the middle, as seen in the magnetic 
field lines shown in the left panel of Figure~\ref{fig:nullimage}. 
The domain has $x\in[-0.5,0.5]$ and $y\in[0,2]$ and is resolved by a uniform mesh of 
$128\times256$ cells. We take $x_c=0$ and $y_c=1$ to make the null point appear at the center
of the domain. To achieve a static equilibrium, the non-zero Lorentz force is balanced by a 
pressure gradient force as
\begin{gather}
 \mathbf{J}\times\mathbf{B}-\nabla p=0 \,.
\end{gather} 
By solving this, we have the gas pressure as $p(x,y)=B_c j_z A_z + p_0$. 
The initial density is uniformly 1, and all is static. The arrows in the right panel of 
Figure~\ref{fig:nullimage} only show the direction of the Lorentz force. We choose dimensionless
values of $B_c=10$, $p_0=7$, $j_t=2.5$, and $j_z=1$. The Alfv\'{e}n crossing time along $y$-axis across
the box is about 1. Plasma $\beta$ ranges from 0.06 at four corners to 8782 near the null point. 
To test how well the numerical solver maintains this initial static equilibrium, we perform runs 
with or without MFS. All variables except for velocity are fixed as their initial values in 
all boundaries, and velocity is made asymmetric about boundary surfaces to ensure zero velocity
 at all boundaries.

After 10 time units (about 22550 CFL limited time steps) in the MFS run, the density 
decreases at most 0.038 \% in the central horizontal line near the side boundaries and increases 
0.02 \% in the central vertical line as shown in Figure~\ref{fig:nullimage}, which corresponds 
to numerically generated diverging and converging flows to the horizontal and vertical line, 
respectively, with very small speeds of magnitude at most $2\times10^{-4}$, which is only a
$6\times10^{-3}$ fraction of the local Alfv\'{e}n speed. This cross-shaped 
pattern has only four cells in width. The reference run without MFS gives almost the same images as in 
Figure~\ref{fig:nullimage}. The MFS run splits off the entire field above as $\mathbf{B_0}$ 
with its non-zero Lorentz force, and gives essentially the same results as the standard 
solver for the equilibrium state. 
\begin{figure}
\includegraphics[width=\textwidth]{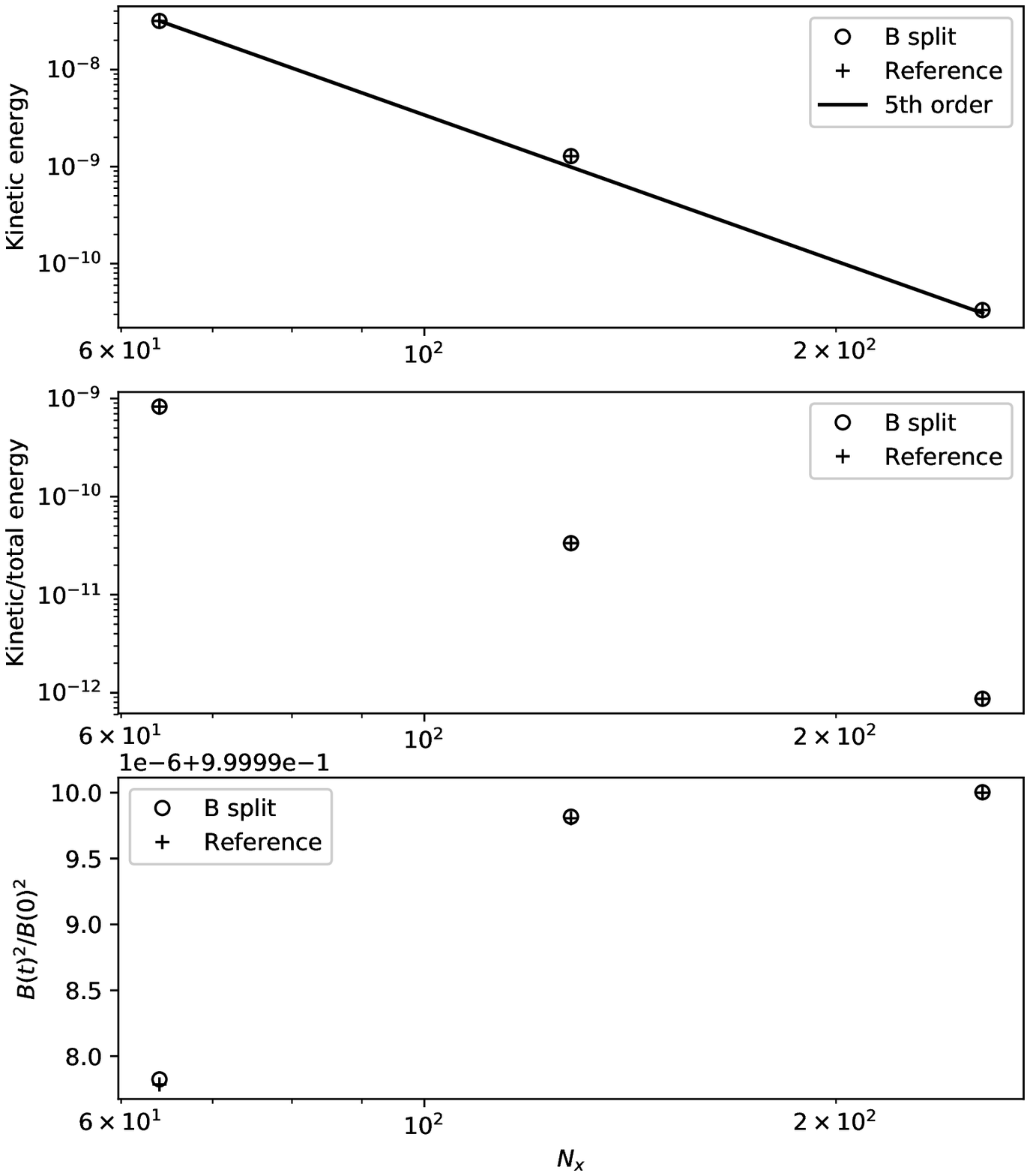}
\caption{Kinetic energy (top panel), the ratio between kinetic energy and total energy (middle panel),
and magnetic energy normalized by its initial value at 10 time units for MFS runs (in circles) 
and reference runs (in plus) with different grid resolutions $64\times128, 128\times256, 256\times512$. 
Note the scale factor for the vertical axis in the bottom panel.}
\label{fig:nullchange}
\end{figure}
We then perform convergence study with increasing grid resolution, and plot kinetic energy, 
the ratio between kinetic energy and total energy, and magnetic energy normalized by its initial 
value at 10 time units in Figure~\ref{fig:nullchange}. Both MFS runs and reference runs without MFS
present very similar converging behavior and errors (deviations from the initial equilibrium) 
decreases with increasing resolution at 5th order of accuracy.
Therefore the correctness of MFS technique for non-force-free fields is justified.

\subsubsection{current sheet diffusion}
To test the MFS technique in resistive MHD, we first test dissipation of a straight current sheet 
with uniform resistivity. The magnetic field changes direction smoothly in 
the current sheet located at $x=0$ as given by
\begin{gather}
 B_x=0 \,, \notag\\
 B_y=-B_d \tanh(c_w x) \,, \notag\\
 B_z= B_d /\cosh(c_w x) \,. \label{eq27}
\end{gather}
Its current density is
\begin{gather}
 J_x=0, \notag\\
 J_y= 5 B_d \tanh(c_w x)/\cosh(c_w x), \notag\\
 J_z=-5 B_d /\cosh^2(c_w x).
\end{gather}
In this test, we adopt parameter values $B_d=4$ and current sheet width $c_w=5$.  It is a non-linear 
force-free field from a previous numerical study of magnetic reconnection of solar flares \citep{Yokoyama01}. 
Initial density and gas pressure are uniformly 1. 

\begin{figure}
\includegraphics[width=\textwidth]{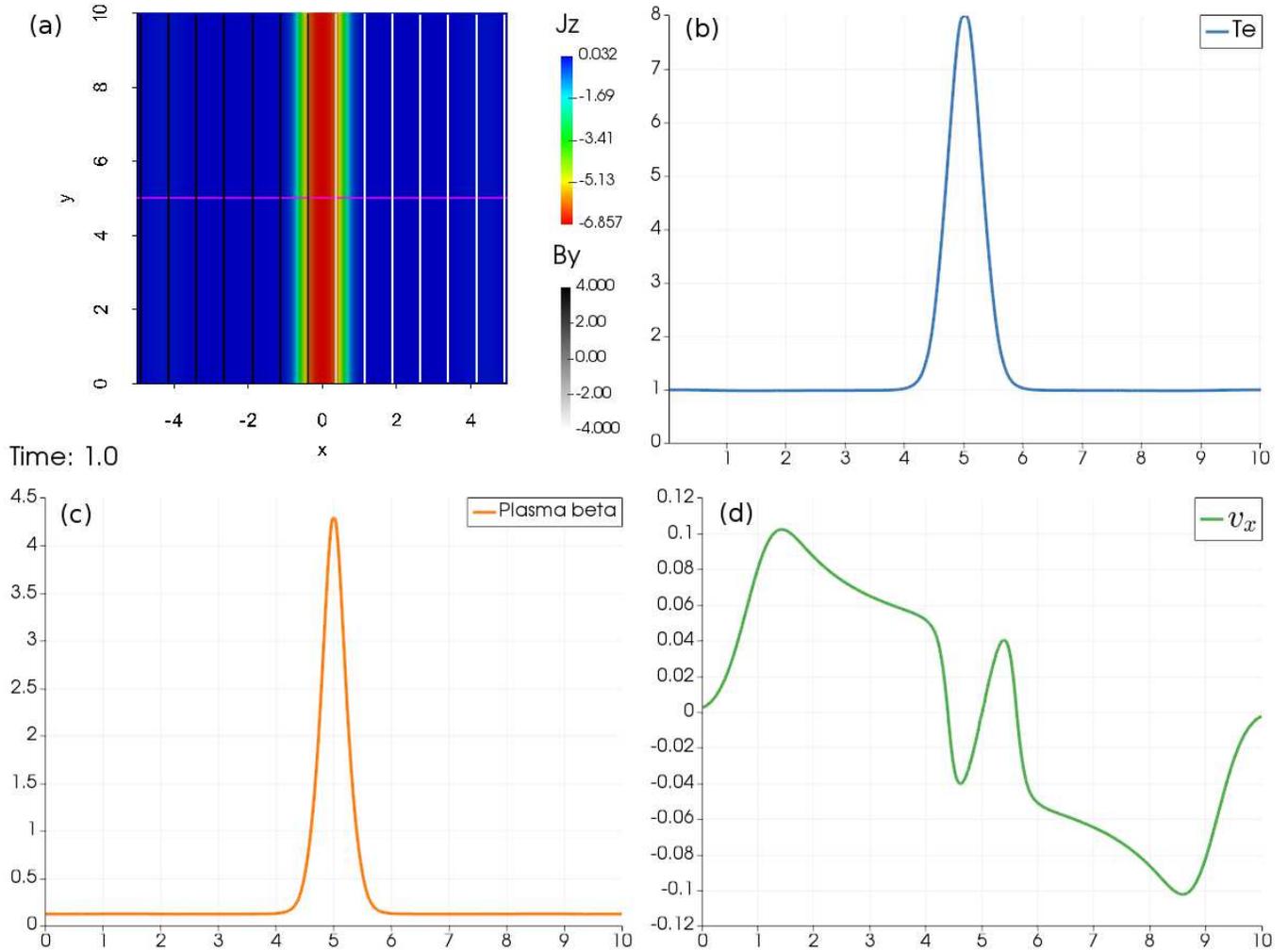}
\caption{Dissipation of current sheet with MFS at 1 time unit. (a) $z$-component of electric 
current over-plotted with magnetic field lines colored by $B_y$. A purple line slicing across 
the current sheet is plotted with its temperature in (b), plasma $\beta$ in (c), and $v_x$ in
(d).}
\label{fig:current}
\end{figure}

We adopt a constant resistivity $\eta$ with a dimensionless value of 0.1. 
The simulation box is 10 by 10 discretized on a 4-level AMR mesh with an effective resolution 
of $512\times512$ cells. The average Alfv\'{e}n crossing time is 2.5. After 1 time unit 
in a MFS run, the current is dissipated due to the resistivity as shown in panel (a) from
Figure~\ref{fig:current}, presenting the $z$-component of current in rainbow color and 
magnetic field lines in grey scale. The Ohmic heating increases the temperature in the current 
sheet from 1 to 8, as shown in panel (b), with flows towards the current sheet from outside
and velocity pointing outward in the heated current sheet as shown in panel (d). Plasma $\beta$ 
ranges from 0.125 outside the current sheet and $4.2$ in the middle of the current sheet.
The reference run without MFS gives undistinguishable images as in Figure~\ref{fig:current}.
\begin{figure}
\includegraphics[width=\textwidth]{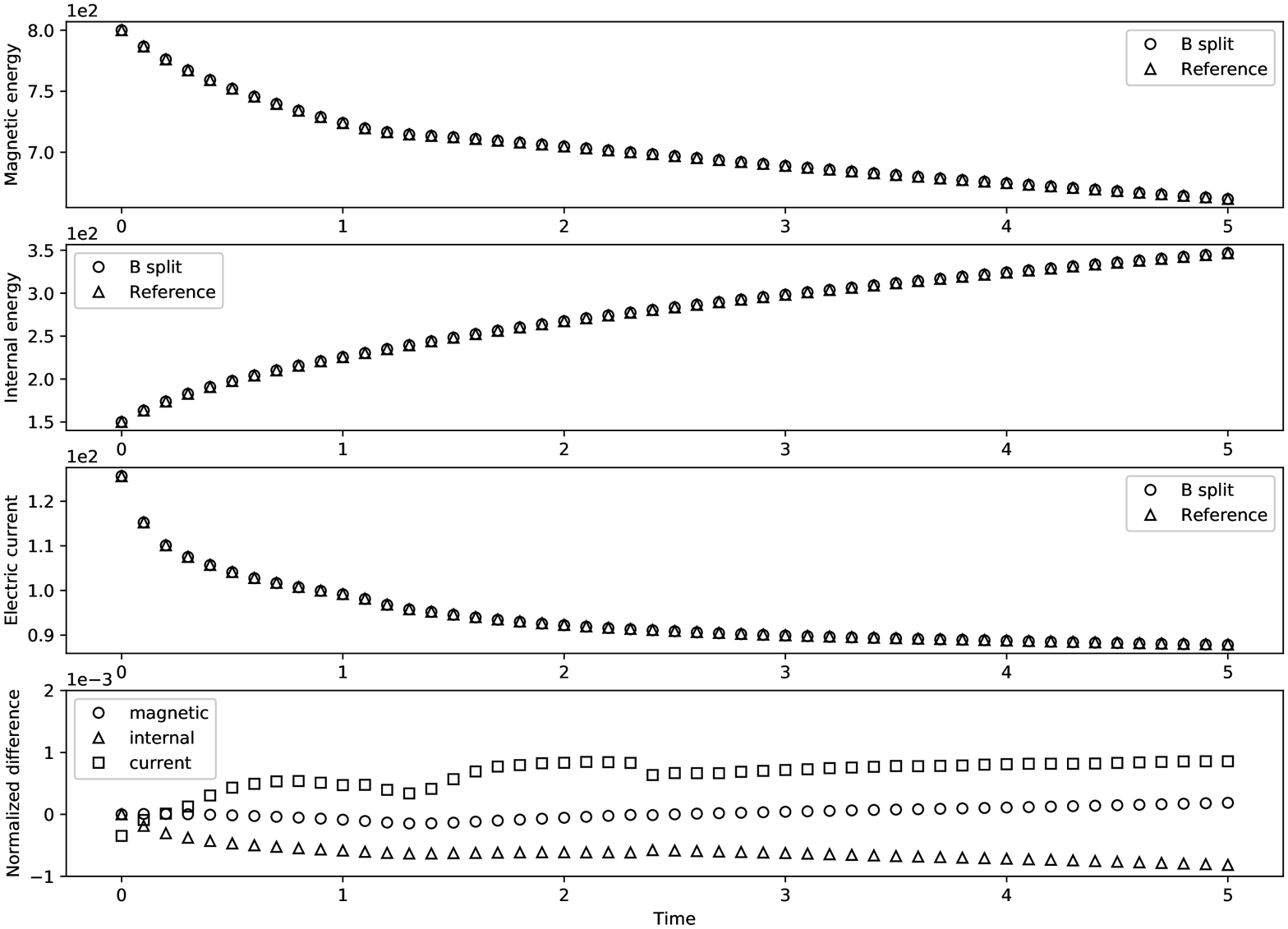}
\caption{Time evolution of (from top to bottom) magnetic energy, internal energy, and electric 
current in the case of MFS (in circles) and the non-split reference case (in triangles). The 
very bottom panel quantifies their relative differences. Note the scale factor in the left top 
corner at each panel.} 
\label{fig:cchange}
\end{figure}
To exactly compare the two runs, we evaluated the time evolution of magnetic energy, internal energy, 
and electric current and plot them in circles for MFS run and in triangles for the reference
run in Figure~\ref{fig:cchange}. The results for the two runs almost overlap and the decrease of magnetic
energy corresponds well to the increase of internal energy, this time by means of the adopted physical dissipation. We plot the normalized 
difference between them in the bottom panel, where within 0.1\% differences are found for all
three quantities. Therefore, the correctness of MFS technique for resistive MHD is verified. 

\subsubsection{Solar flare application with anomalous resistivity}
To show the advantage of MFS technique in fully resistive settings, we present its application 
on a magnetic reconnection test which extends the simulations of solar flares by \citet{Yokoyama01} 
to extremely low-$\beta$, more realistic situations. Here, a run without the MFS technique 
using our standard MHD solver fails due to negative pressure. 

We setup the simulation box in $x\in[-10,10]$ and $y\in[0,20]$ with a four-level AMR mesh, which
has an effective resolution of $2048\times2048$ and smallest cell size of 97 km.
The magnetic field setup is the same current sheet as before, and follows 
equation~(\ref{eq27}) with parameters $B_d=50$ and $c_w=6.667$. Now, a 2D setup is realized 
with vertically varying initial density and temperature profiles as in~\citet{Yokoyama01}, 
where a hyperbolic tangent function was used to smoothly connect a corona region with a 
high-density chromosphere. By neglecting solar gravity and set constant gas pressure 
everywhere, the initial state is in an equilibrium. The current sheet is then first
perturbed by an anomalous resistivity $\eta_1$ in a small circular region in the current sheet when
time $t<0.4$, and afterwards the anomalous resistivity $\eta_2$ is switched to a more 
physics-based value depending on the relative ion-electron drift velocity. The formula of
anomalous resistivity $\eta_1$ is 
\begin{equation}
\eta_1=\begin{cases} \eta_c [2(r/r_\eta)^3-3(r/r_\eta)^2+1]& \text{if } r \le r_\eta, \\
               0 & \text{if } r > r_\eta,
  \end{cases}
\end{equation}
where $\eta_c=0.002$ is the amplitude, $r=\sqrt{x^2+(y-6)^2}$ is the distance from the 
center of the spot at $(x=0, y=6)$, and $r_\eta=0.24$ is the radius of the spot.
The formula of anomalous resistivity $\eta_2$ is
\begin{equation}
\eta_2=\begin{cases} 0 & \text{if } v_d < v_c, \\
               \alpha_\eta (v_d/v_c-1) & \text{if } v_d \ge v_c.
  \end{cases}
\end{equation}
where $v_d=\sqrt{J_x^2+J_y^2+J_z^2}/(en)$ is the relative ion-electron drift velocity
($e$ is the electric charge of electron and $n$ is number density of plasma), 
$v_c=5\times10^{-5}$ is the threshold of the anomalous resistivity, and $\alpha_\eta=4\times10^{-3}$
 is the amplitude.

Continuous boundary conditions are applied for all boundaries and all variables except that $B_x=0$
and zero velocity is imposed at the bottom boundary. We basically adopt the same setup and physical 
parameters as in~\citet{Yokoyama01}, except that the magnitude of magnetic field here reaches 100 G with plasma 
$\beta$ of 0.005, compared to about 76 G and $0.03$ in the original work. We use normalization units of 
length $L_u=10$ Mm, time $t_u=85.87$ s, temperature $T_u=10^6$ K, number density $n_u=10^9$ cm$^{-3}$, 
velocity $v_u=116.45$ km s$^{-1}$, and magnetic field $B_u=2$ G. Since we use different 
normalization units to normalize the same physical parameters, our dimensionless values are 
different from Yokoyama's work. We choose a scheme an combining HLL flux with van Leer's slope limiter 
\citep{vanLeer74} and two-step time integration. Powell's eight-wave method \citep{Powell99} is used to 
clean divergence of magnetic field. Thermal conduction along magnetic field lines is solved using techniques 
 that are described in details in the following Section~\ref{sec:conduction}.

\begin{figure}
\includegraphics[width=\textwidth]{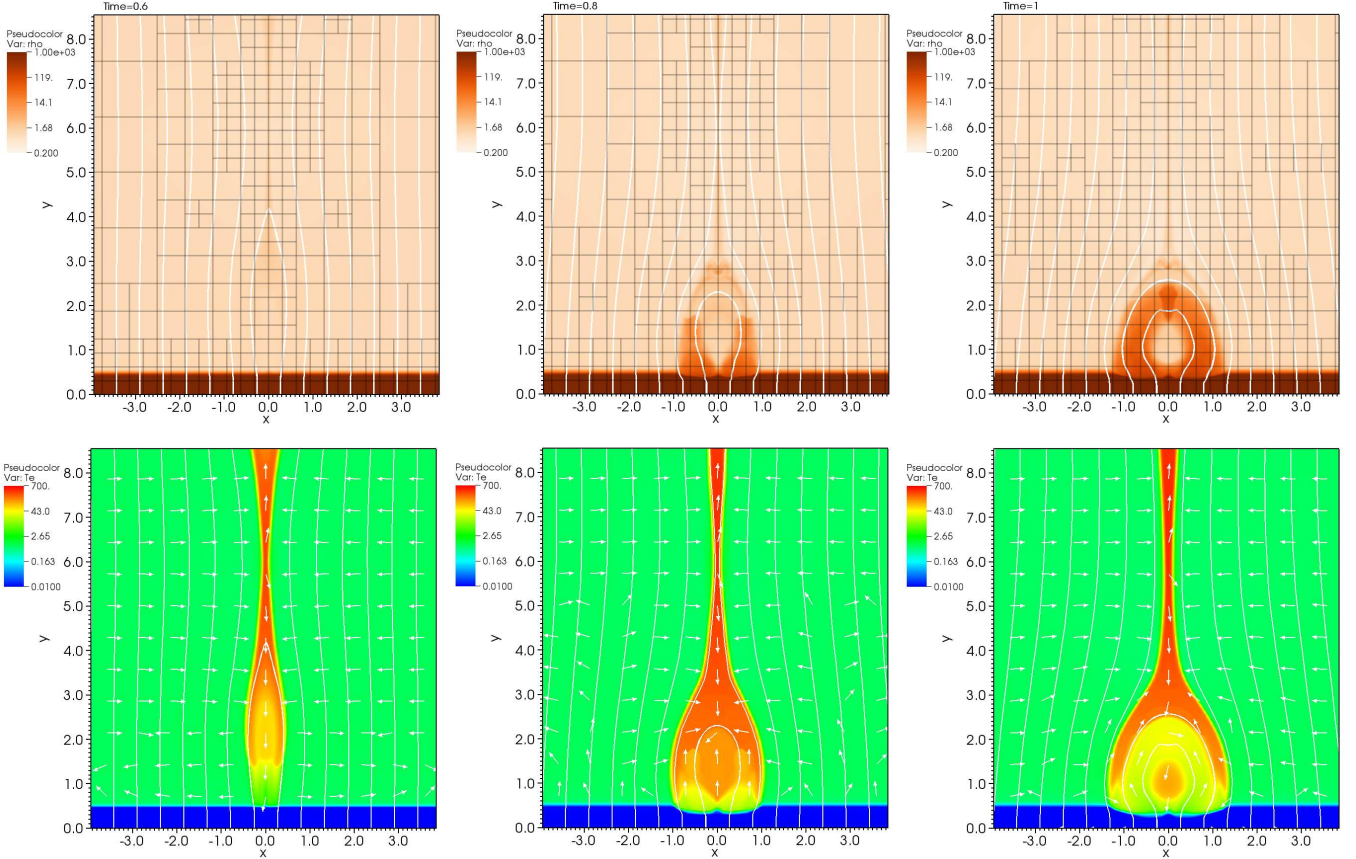}
\caption{Results of the solar flare simulation. Temporal evolution of density (top) and temperature 
(bottom) distribution. The arrows show the directions of velocity, and lines show the magnetic 
field lines integrated from fixed seed points at the bottom. Boarders of AMR blocks are plotted
in the density maps. Values in color bars are dimensionless.} 
\label{fig:flare}
\end{figure}
Inspired by Figure~3 in~\citet{Yokoyama01}, we show very similar results in 
Figure~\ref{fig:flare} which presents snapshots of the density and temperature distribution at
 times 0.5, 0.64, and 0.75. The temporal evolution of the magnetic reconnection process and the 
evaporation in flare loops are reproduced.
Flow directions and magnetic field lines are also plotted. Magnetic reconnection starts at the 
center $x=0$, $y=6$ of the enhanced resistive region, where an X-point is formed. Reconnected field 
lines with heated plasma are continuously ejected from the X-point to the positive and negative
$y$-directions because of the tension force of the reconnected field lines.

\section{Anisotropic thermal conduction}\label{sec:conduction}

Thermal conduction in a magnetized plasma is anisotropic with respect to the magnetic field, 
as the thermal conductivity parallel to magnetic field lines is much larger than the one 
perpendicular to field lines. This dominantly field-aligned heat flow can be represented by 
adding the divergence of an anisotropic heat flux to the energy equation. This heat source 
term can be numerically solved, independently of the MHD equations, using an operator splitting 
strategy. This means that we ultimately face the challenge to solve an equation due to 
anisotropic thermal conduction as:
\begin{gather}
\frac{\partial e}{\partial t}=\nabla\cdot\mathbf{q} \,,  \label{heat}\\
\mathbf{q}=(\kappa_\parallel-\kappa_\perp) (\mathbf{b}\cdot\nabla T) \mathbf{b}+\kappa_\perp\nabla T \,,
\end{gather}
where $e$ is the internal energy $e=p/(\gamma-1)$ per unit volume, or it is the total energy 
density $E$ from before. In this equation, $T$ is temperature, $\mathbf{q}$ 
is the (negative) heat flux vector, $\mathbf{b}=\mathbf{B}/B=(b_x,b_y,b_z)$ is the unit vector along the magnetic 
field. The term $\mathbf{b}\cdot\nabla T$ represents the temperature gradient along the magnetic field. 
$\kappa_\parallel$ and $\kappa_\perp$ are conductivity coefficients for parallel and 
perpendicular conduction with respect to the local field direction. In most magnetized 
astronomical plasma, $\kappa_\parallel$ is dominating, for example in the solar corona, where $\kappa_\perp$ is 
about 12 orders of magnitude smaller than $\kappa_\parallel$. Therefore, we can typically ignore physical 
perpendicular thermal conduction, but we nevertheless implemented 
both parallel and perpendicular thermal conduction for completeness. Note than when 
$\kappa_\perp=\kappa_\parallel$, the thermal conduction naturally degenerates 
to isotropic one. In typical astronomical applications, we typically choose the Spitzer conductivity for 
$\kappa_\parallel=8\times10^{-7}T^{5/2}$ erg cm$^{-1}$ s$^{-1}$ K$^{-1}$. 

\subsection{Discretiziting the heat conduction equation}
To solve the thermal conduction equation~(\ref{heat}), various numerical schemes
based on centered differencing have been developed. The simplest example is to
evaluate gradients directly with second order central differencing; more
advanced methods include the asymmetric scheme described
in~\citep{Parrish05,Gunter05} and the symmetric scheme of \citep{Gunter05}.
However, these schemes can cause heat to flow low to high temperature if there
are strong temperature gradients~\citep{Sharma07}, which is unphysical.

To understand why such effects can occur, consider equation \eqref{heat}.
Suppose a finite volume discretization is used in 2D, and that we want to compute 
the heat flux in the $x$-direction at the cell interface between cell $(i,j)$ and 
$(i+1,j)$. The term
$(\kappa_\parallel-\kappa_\perp)(\mathbf{b}\cdot\nabla T)\mathbf{b}$ is then
given by
$(\kappa_\parallel-\kappa_\perp)(b_x^2 \partial_x T + b_x b_y \partial_y T)$.
The term $b_x^2 \partial_x T$ will always be heat flux transported from hot to cold.
When $b_x b_y \partial_y T$ has the opposite sign of $b_x^2 \partial_x T$ and a larger
magnitude, a resulted unphysical heat flux from cold to hot occurs.

Avoiding the above problem, which can lead to negative temperatures, is
especially important for astrophysical plasmas with sharp temperature gradients.
Examples are the transition region in the solar corona separating the hot corona
and the cool chromosphere, or the disk-corona interface in accretion disks. A
solution was proposed in \citet{Sharma07}, in which slope limiters were applied
to the transverse component of the heat flux. 
Since the slope limited symmetric scheme was found to be less diffusive than the limited asymmetric 
scheme, we follow \citet{Sharma07} and implemented it in our code, with the extra extension to 
(temperature dependent) Spitzer conductivity and generalized to a 3D setup in Cartesian coordinates.
The explicit, conservative finite difference formulation of equation~(\ref{heat}) is then
\begin{gather}
e^{n+1}_{i,j,k}=e^n_{i,j,k}+\Delta t\left(\frac{q^n_{x,i+1/2,j,k}-q^n_{x,i-1/2,j,k}}{\Delta x}+\right. \\
                            \left.\frac{q^n_{y,i,j+1/2,k}-q^n_{y,i,j-1/2,k}}{\Delta y}+
                            \frac{q^n_{z,i,j,k+1/2}-q^n_{y,i,j,k-1/2}}{\Delta z}\right) \,,
\end{gather}
involving face-centered evaluations of heat flux, e.g. along the $x$-coordinate given by
\begin{gather}
q_x=\kappa b_x\left(b_x\frac{\partial T}{\partial x}+b_y\frac{\partial T}{\partial y}+ 
b_z\frac{\partial T}{\partial z}\right)+\kappa_\perp\frac{\partial T}{\partial x}, \label{eq:qx}
\end{gather}
in which $\kappa=\kappa_\parallel-\kappa_\perp$. 

Since we use an explicit scheme for time integration, the time step $\Delta t$ must
satisfy the condition for numerical stability:
\begin{equation}
\Delta t\leqslant c_p \min\left(\frac{\rho \Delta x^2}{(\gamma-1)\kappa_\parallel b_x^2}, 
\frac{\rho \Delta y^2}{(\gamma-1)\kappa_\parallel b_y^2},\frac{\rho \Delta z^2}{(\gamma-1)\kappa_\parallel b_z^2}\right),
\end{equation}
where $c_p$ is a stability constant taking value of 0.5 and 1/3 for 2D and 3D, respectively.
With increasing grid resolution and/or high thermal conductivity, this limit on $\Delta t$ can be much
smaller than the CFL time step limit for solving the ideal MHD equations explicitly. Therefore multiple steps of 
solving thermal conduction within one step MHD evolution are suggested to alleviate the
time step restriction from thermal conduction. In our actual solar coronal applications, we adopt a super 
timestepping RKL2 scheme proposed by \citet{Meyer12}, where every parabolic update uses an $s$-stage 
Runge-Kutta scheme with $s$ and its coefficients determined in accord with the 2-term recursion formula 
for Legendre polynomials. In what follows, we just concentrate on the details to handle a single 
step for the thermal conduction update. 

The first term on the right hand side of expression~(\ref{eq:qx})
is discretized as the limited face-centered heat flux:
\begin{equation}
q_{xx,i+1/2,j,k}=\kappa_{i+1/2,j,k}\left(\overline{b_x^2\frac{\partial T}{\partial x}}\right)_{i+1/2,j,k},
\end{equation}
where the local thermal conduction coefficient is found by first quantifying its value at the cell corners, in 3D ($n_{\rm D}=3$) found from 
\begin{gather}
\kappa_{i+1/2,j+1/2,k+1/2}=\frac{1}{2^{n_{\rm D}}}\sum^1_{n=0}\sum^1_{m=0}\sum^1_{l=0}\kappa_{i+l,j+m,k+n} \,, \label{eq:kcorn}
\end{gather}
to then derive the face-centered values as in
\begin{gather}
\kappa_{i+1/2,j,k}=\frac{1}{2^{n_{\rm D}-1}}\sum^1_{n=0}\sum^1_{m=0}\kappa_{i+1/2,j+1/2-m,k+1/2-n} \,. \label{eq:kface}
\end{gather}
The face-centered, limited flux expression is given by
\begin{equation}
\left(\overline{b_x^2\frac{\partial T}{\partial x}}\right)_{i+1/2,j,k}=\frac{1}{2^{n_{\rm D}-1}}\sum^1_{n=0}\sum^1_{m=0}
  \left(b_x^2\widetilde{\frac{\partial T}{\partial x}}\right)_{i+1/2,j+1/2-m,k+1/2-n}, \label{eq:qnorm}\\
\end{equation}
in which corner values for the magnetic field aligned unit vectors are found as in
\begin{gather}
b_{x,i+1/2,j+1/2,k+1/2}=\frac{1}{2^{n_{\rm D}}}\sum^1_{n=0}\sum^1_{m=0}\sum^1_{l=0}b_{x,i+l,j+m,k+n}\,, 
\end{gather}
while the limited temperature gradient in cell corners is found from
\begin{gather}
\left.\widetilde{\frac{\partial T}{\partial x}}\right\rvert_{i+1/2,j+1/2-m,k+1/2-n}=\begin{cases} 
\bar{q} & \text{if } 
\min q < \bar{q} < \max q, \\ 
\min q & \text{if } \bar{q} \leqslant \min q, \\
\max q & \text{if } \bar{q} \geqslant \max q, \\
\end{cases}\\
\text{where } \bar{q}=\left.\overline{\frac{\partial T}{\partial x}}\right\rvert_{i+1/2,j+1/2-m,k+1/2-n} \,,\\
\left.\overline{\frac{\partial T}{\partial x}}\right\rvert_{i+1/2,j+1/2,k+1/2}=\frac{1}{2^{n_{\rm D}-1}}
\sum^1_{a=0}\sum^1_{b=0}\left.\frac{\partial T}{\partial x}\right\rvert_{i+1/2,j+a,k+b}, \\
\min q=\min(\alpha \left.\frac{\partial T}{\partial x}\right\rvert_{i+1/2,j,k}, 1/\alpha \left.\frac{\partial T}{\partial x}\right\rvert_{i+1/2,j,k}), \\
\max q=\max(\alpha \left.\frac{\partial T}{\partial x}\right\rvert_{i+1/2,j,k}, 1/\alpha \left.\frac{\partial T}{\partial x}\right\rvert_{i+1/2,j,k}), \\
\text{and finally }\left.\frac{\partial T}{\partial x}\right\rvert_{i+1/2,j,k}=(T_{i+1,j,k}-T_{i,j,k})/\Delta x. 
\end{gather}
Similar formulae (involving less summations in the averaging) apply at lower dimensionality $n_{\rm D}$. 
We choose $\alpha=0.75$ for 2D runs, the same value as used by \citet{Sharma07}, and $\alpha=0.85$ for 3D
runs to achieve least numerical diffusion according to our try-and-error numerical tests.
Note that in our code, as assumed in the above, the magnetic field components (and other quantities, 
like temperature) are defined at cell centers, and different formulae would be needed for a staggered grid
where magnetic field is defined at cell faces, such as commonly adopted in constrained transport schemes.

The second and the third term of equation~(\ref{eq:qx}) are transverse components $q_{xy}$ and $q_{xz}$
of heat conduction, and we illustrate the discretization of the second term
here and the third term can be derived similarly:
\begin{equation}
q_{xy,i+1/2,j,k}=\kappa_{i+1/2,j,k}\left(b_xb_y\overline{\frac{\partial T}{\partial y}}\right)_{i+1/2,j,k},
\end{equation}
where $b_{x,i+1/2,j,k}=0.5(b_{x,i,j,k}+b_{x,i+1,j,k})$ and $b_{y,i+1/2,j,k}$ follow the same formula as 
equation (\ref{eq:kface}) for $\kappa_{i+1/2,j,k}$. The limited, face-centered temperature gradient is evaluated as in
\begin{equation}
\left.\overline{\frac{\partial T}{\partial y}}\right\rvert_{i+1/2,j,k}=
L\left(\left.\frac{\partial T}{\partial y}\right\rvert^L_{i,j,k},
\left.\frac{\partial T}{\partial y}\right\rvert^L_{i+1,j,k}
\right)
\end{equation}
in which
\begin{equation}
\left.\frac{\partial T}{\partial y}\right\rvert^L_{i,j,k}=L\left(\left.\frac{\partial T}{\partial y}\right\rvert_{i,j+1/2,k},
\left.\frac{\partial T}{\partial y}\right\rvert_{i,j-1/2,k}\right)
\end{equation}
where $L(a,b)$ is the monotonized central (MC) slope limiter, which is given by 
\begin{equation}
{\rm MC}(a,b)=2\sign(a)\max\{0,\min[|a|,\sign(a)b,\sign(a)(a+b)/4]\}.
\end{equation}
We have tried other slope limiters such as minmod, vanleer, and superbee, but 
MC limiter gives the best results.

The last term of equation~(\ref{eq:qx}) contributes to perpendicular thermal 
conduction and its discretized form computes the averaged face value of 
$\kappa_{\perp,i+1/2,j,k}$ using the same form as equation~(\ref{eq:kface}) 
and the averaged face value of the temperature gradient along $x$ as in 
equation~(\ref{eq:qnorm}) (and the equations that follow it), but without the factors $b_x^2$.

Since thermal conduction in dilute plasmas is caused by the movement of electrons, a
larger thermal conduction flux implies faster electrons. Once the electron speed becomes 
comparable to the sound speed, plasma instabilities prevent the electrons from flowing 
faster and thermal conduction must saturate~\citep{Cowie77}. We adopt the following formula from \citet{Cowie77} for a saturated thermal flux:
\begin{equation}
q_{\rm sat}=5\phi\rho c_{\rm iso}^3\,,
\end{equation}
where $c_{\rm iso}$ is the isothermal sound speed and the factor $\phi=1.1$ when the 
electron and ion temperatures are equal. We evaluate the projected saturated thermal flux at
each cell face based on the averaged density and temperature there, and set it as the 
upper limit for the total thermal flux through that cell face, namely, 
$|q_{x,i+1/2,j,k}|\leqslant q_{\text{sat},i+1/2,j,k} |b_{x,i+1/2,j,k}|$. 

We implement the methods mentioned above in a thermal conduction module with source 
code written in the dimension-independent \texttt{LASY} way.
Thus the same compact source code can be translated to pure \texttt{Fortran} code in 1D, 2D, or 3D
before compilation.

\subsection{Anisotropic heat conduction test problems: 2D to 3D}
A 2D circular ring diffusion test problem was proposed by \citep{Parrish05} and became a 
standard test for anisotropic thermal conduction \citep{Sharma07,Meyer12}. It 
initializes a hot patch in circular magnetic field lines in a $[-1,1]\times[-1,1]$ 
Cartesian box with temperatures
\begin{equation}
T=\begin{cases}12 & \text{if } 0.5 < r < 0.7 \text{ and } \frac{11}{12}\pi < \theta < \frac{13}{12} \pi, \\
               10 & \text{otherwise},
  \end{cases}
\end{equation}
where $r=\sqrt{x^2+y^2}$ and $\tan\theta=y/x$. Density is set to 1. We create
circular magnetic field lines centered at the origin by the formula:
\begin{equation}
B_x=10^{-5}\cos(\theta+\pi/2)/r,~~~
B_y=10^{-5}\sin(\theta+\pi/2)/r,
\end{equation}
which is generated by a infinite line current perpendicular to the plane through the
origin. The parallel thermal conductivity is constant and set to $\kappa_\parallel=0.01$ and no explicit 
perpendicular thermal conduction is considered. Continuous boundary condition is used
for all variable at all boundaries. We only evolve the anisotropic 
thermal heat conduction equation without solving the MHD equations, until 400 time units. 
Ideally, the diffusion of temperature should only occur in a ring $0.5 < r < 0.7 $ and
eventually result in a uniform temperature of $10.1667$ inside the ring with 
unperturbed temperature of 10 outside the ring. But perpendicular numerical diffusion 
may cause some cross-field thermal conduction and can smooth sharp temperature gradients
at borders of this ring.

\begin{figure}
\includegraphics[width=\textwidth]{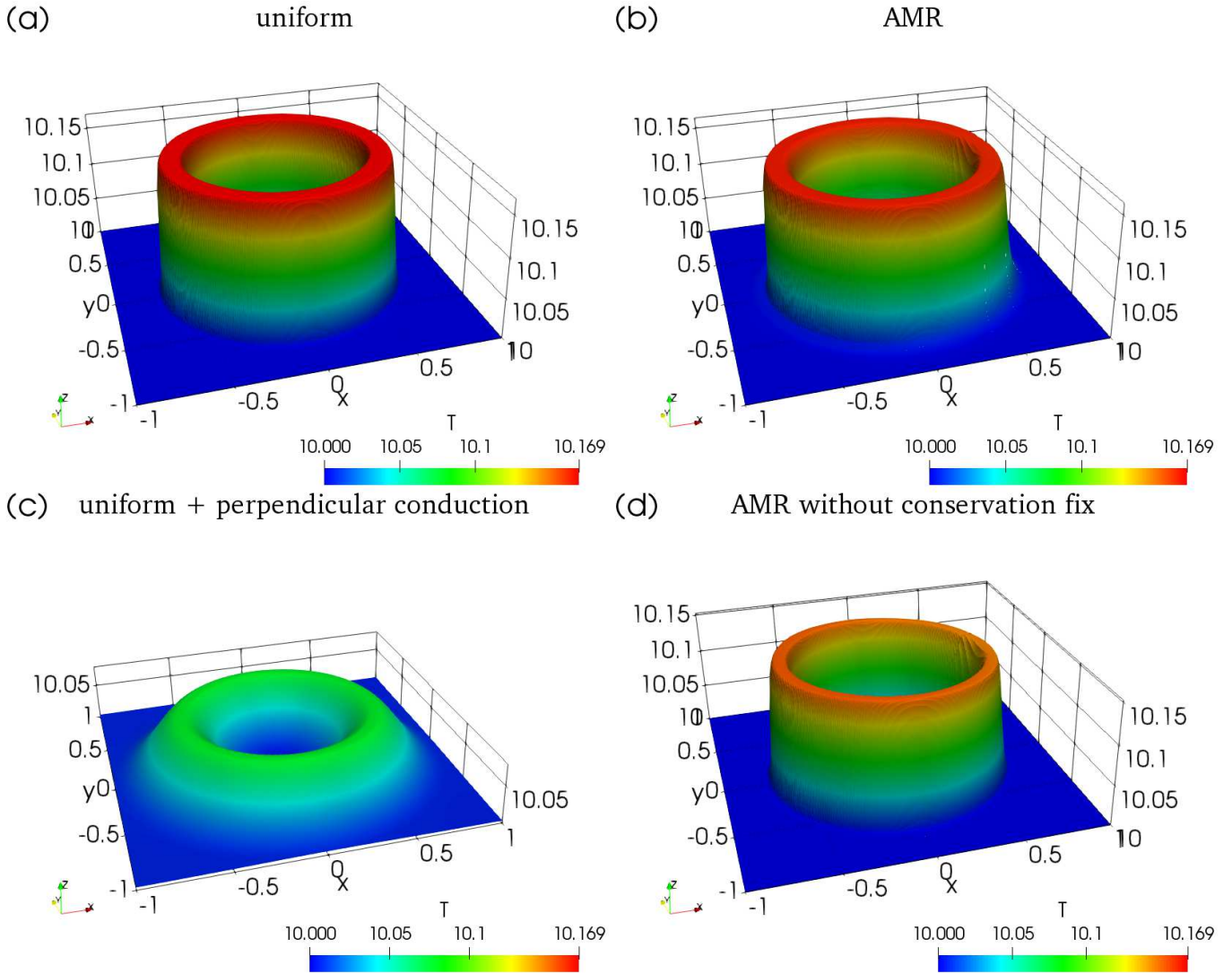}
\caption{Temperature at 400 time units for four cases initialized with the ring 
diffusion problem. (a) on uniform $200\times200$ grid; (b) on AMR grid with 
$200\times200$ effective resolution; (c) the same condition as (a) except that 
perpendicular conduction with $\kappa_\perp=0.00005$ is included; (d) the same condition 
as (b) except for excluding conservation fix.}
\label{fig:ringtc}
\end{figure}
The end state of a run (a) on a uniform $200\times200$ grid is shown in 
Figure~\ref{fig:ringtc}(a). Temperature is high and nearly uniform inside the ring, with 
a constant background temperature of 10. The maximal and minimal temperature of 
the end state are 10.1687 and 10, respectively. No temperature overshooting is found near
sharp temperature gradient regions and monotonicity property is preserved. 
The final temperature distribution agrees with previous findings \citep[see the fourth 
panel in Figure 6 of][]{Sharma07}. Since the exact solution of this problem is a uniform temperature
of $10.1667$ in the ring and 10 outside it, we can evaluate the $L_1$, $L_2$, and $L_\infty$
norm errors for four runs with increasing resolutions (see table~\ref{tab:tcam}). 
The $L_1$ norm converges with a
rate of 1.4 and $L_2$ norm converges with a rate of 0.8. The maximal temperature is
getting closer to the correct solution with increasing resolution and the minimal 
temperature stays at the correct value of 10. A rough estimate for the time averaged 
perpendicular numerical diffusion (conductivity) $\kappa_{\perp,\rm{num}}$ and its ratio
to $\kappa_\parallel=0.01$ are calculated 
following the same method in \citet{Sharma07}, and reported in the last column of 
table~\ref{tab:tcam}. Note in particular that we can get this ratio from order 
$10^{-3}$ down to order $10^{-5}$ on sufficiently large grids, so that we can add 
physical perpendicular thermal conduction coefficients that exceed these values.
\begin{table}
\begin{center}
\caption{Accuracy measures of the ring diffusion tests on uniform grids\label{tab:tcam}}
\begin{tabular}{ccccccc}
\hline
Resolution & $L_1$ error & $L_2$ error & $L_\infty$ error & $T_{\rm max}$ & $T_{\rm min}$ 
& $\kappa_{\perp,\rm{num}}/\kappa_\parallel$\\
\hline
$ 50\times 50$  & 0.03037 & 0.04705 & 0.08617 & 10.0842 & 10 & $4.345\times10^{-3}$\\
$100\times100$  & 0.01338 & 0.02704 & 0.11654 & 10.1355 & 10 & $5.145\times10^{-4}$\\
$200\times200$  & 0.00521 & 0.01592 & 0.08683 & 10.1663 & 10 & $1.050\times10^{-4}$\\
$400\times400$  & 0.00320 & 0.01224 & 0.08758 & 10.1681 & 10 & $3.353\times10^{-5}$\\
\hline
\end{tabular}
\end{center}
\end{table}

In order to use the thermal conduction on a AMR grid, refluxing operations 
are needed to ensure flux conservation across fine-coarse interfaces. We
store the thermal fluxes of blocks with finer or coarser neighbor blocks, and send 
thermal fluxes at fine-coarse interfaces from finer blocks to coarser blocks. This is 
standard practice for ensuring conservation in finite volume AMR computations, and our 
strategy in terms of parallel communication patterns involved is found in section 3.4 
of~\citet{Keppens12}. Figure~\ref{fig:ringtc}(b) shows the end state of run (b) on an 
AMR grid with three levels and an effective resolution of $200\times200$. We use 
dynamic AMR, with the refine-triggering based on total energy using a L\"{o}hner type estimator 
\citep{Lohner87, Keppens12} and set a refine threshold of 0.08. Comparing to the 
result of the run in panel (a) on a uniform grid with the same resolution, run (b) 
gives comparable results. The AMR advantage is in the fact that it uses  about 20\% less 
computational time. Some minor quality degeneracy is present, with less sharp boarders 
of the ring segment at the far end from the initial hot segment and a lower maximal 
temperature of 10.1638. This is likely caused by the larger
numerical diffusion associated with the coarser grids at the far end. To prove the importance of the
conservation fix, we show a run (d) with the same setup as run (b), except that it has 
no conservation fix, which ends up with lower temperature in the ring and stronger 
diffusion at its boarders, as shown in Figure~\ref{fig:ringtc}(d).
To test perpendicular thermal conduction, we perform test run (c) with a constant
$\kappa_\perp=5\times10^{-5}=5\times10^{-3}\kappa_\parallel$ and other setups the 
same as run (a). The final hot ring is significantly smoothed by perpendicular 
thermal conduction.

\begin{figure}
\includegraphics[width=\textwidth]{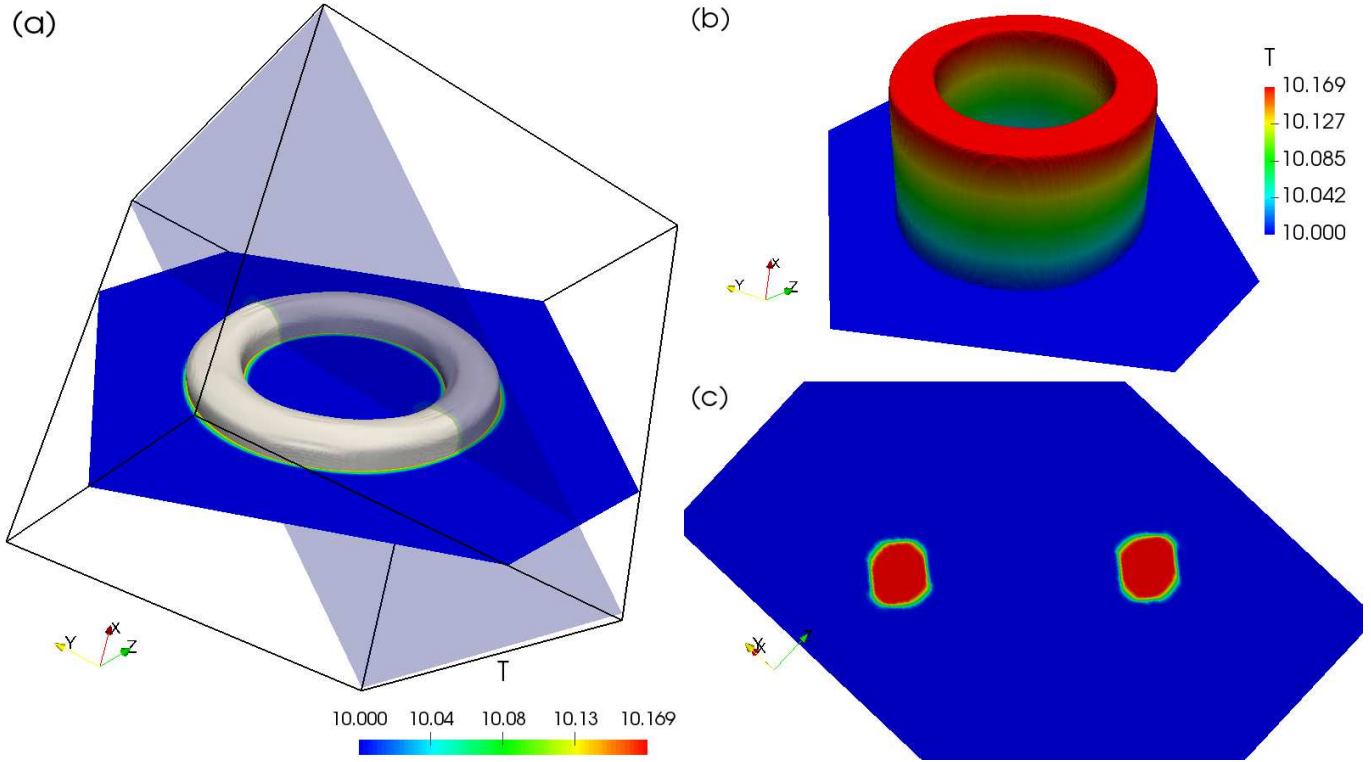}
\caption{Temperature at 400 time units in a 3D case initialized with the ring 
diffusion problem on a uniform $200\times200\times200$ grid. (a) isosurface
of temperature at 10.16 and a plane slicing through the middle of the hot patch in the
3D box; (b) 3D view of the 2D slice plane in (a), warped and colored by temperature;
(c) temperature on the translucent slice in (a).}
\label{fig:ringtc3d}
\end{figure}

We generalize the 2D ring diffusion test to a 3D version. In a 3D $[-1,1]^3$ Cartesian box, 
we rotate the magnetic field loops, from a position at first in $x$-$y$ planes, around $x$-axis 45 degrees and then rotate around 
$z$-axis 45 degrees, to make the magnetic field direction misaligned with the 
coordinates. Therefore, the thermal flux along magnetic field lines must include
contributions from all three dimensions. An initial 3D hot patch is set to have the 
same extension in the plane of the magnetic loop as in the 2D case and an extension 
of $0.4$ in the direction perpendicular to this plane. We run the test on a uniform 
$200\times200\times200$ mesh until 400 time units. As 
shown in Figure~\ref{fig:ringtc3d}(a), the 3D isosurface of temperature at 10.16 
represents a ring shape. A slice through the middle plane of the ring structure is 
shown in a warped view and colored by temperature in Figure~\ref{fig:ringtc3d}(b), and
is comparable with the previous 2D results. The maximal 
temperature in the hot ring of the slice is 10.1774 and the thickness of the hot 
ring is 10\% larger than in the corresponding 2D tests. A translucent slice cutting 
through the ring in panel (a) shows, in panel (c), that its cross section has a 
rounded square shape, which is caused by large numerical diffusion at its corner edges.
To test the 3D thermal conduction on a AMR mesh, we run the same 3D test problem on 
a 3-level AMR mesh with base-level resolution of $50\times50\times50$, which has the
same effective resolution as the previous test on a uniform mesh. The result is very 
similar to the uniform-mesh version with maximal temperature of 10.1727, while the run
consumes 45\% less computational time than the uniform-mesh run.

To check the consistency between the 3D implementation and the 2D one, we set up a test of 2D
ring diffusion problem in a 3D simulation. We set the magnetic field loops in $x$-$y$ planes 
and the initial hot patch has the same extension in $x$ and $y$ directions as in the 2D setup 
and extends uniformly through the box in $z$ direction. Periodic boundary condiction is used
at $z$ boundaries and continuous boundary condiction is adopted in $x$ and $y$ boundaries.
The simulation box has a size of 1 by 1 by 0.1 on a $200\times200\times20$ mesh. The 
temperature distribution is invariant in $z$ direction and its $x$-$y$ plane slice is very 
close to the 2D result with the same resolution at 400 time units. The maximum and miminum 
temperatures are the same. The temperature transition region is about 4\% broader in the 
3D run. Therefore, the 3D implementation is consistent with the 2D one.

\section{Performance and scaling}\label{sec:scaling}

\begin{figure}
\includegraphics[width=\textwidth]{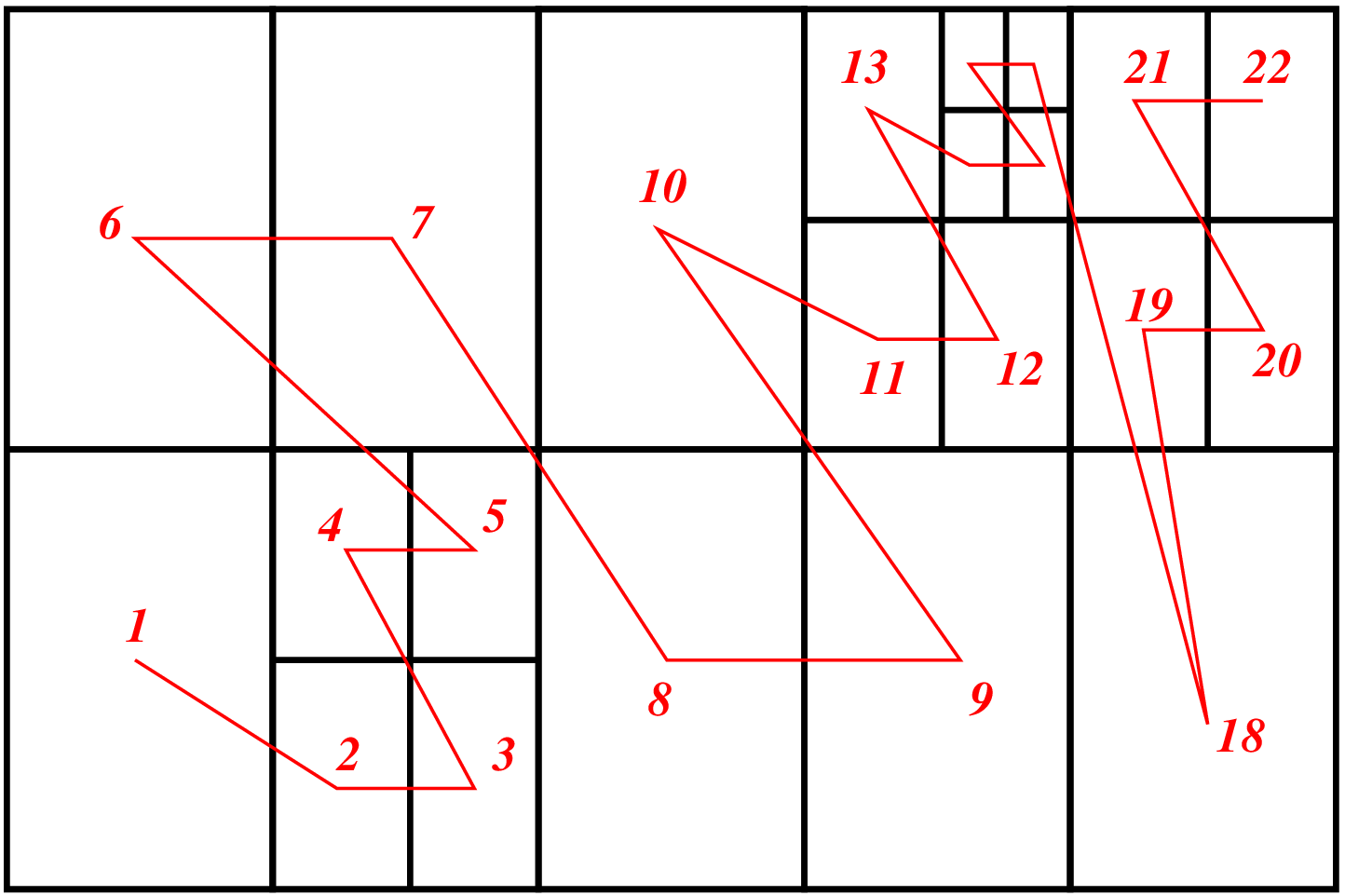}
\caption{Morton ordered curve illustrated for a rectangular domain with $5\times2$ root level blocks.}
\label{Morton}
\end{figure}

In the previous version of \texttt{MPI-AMRVAC}, the root-level blocks were connected with a simple
dimension-by-dimension incremental order curve, while a Morton-ordered space filling 
curve was recursively applied from level 2 onwards, refining root level blocks, to the leaf levels 
\citep{Keppens12}. In \texttt{MPI-AMRVAC 2.0}, we also connect the root-level blocks with a Morton-ordered space filling
curve, as illustrated in Figure~\ref{Morton}, to improve data locality and minimize times of 
inter-processor communications. Since a pure Morton curve connects nodes arranged in a 
square shape in 2D or in a cubic shape in 3D which have the same even number of node blocks 
along each dimension, we can not directly apply it to a domain with nodes arranged in a 
non-square shape or with odd number of node blocks.  
In a true rectangular case, e.g. $5\times2$ in Figure~\ref{Morton}, we first extend the 
rectangular domain to a square one, namely, $6\times6$, apply the standard Morton curve on it, 
and remove nodes that lay outside of the rectangular domain. To remove a node on a space 
filling curve, we simply delete the node and subtract one from the numbers of 
all nodes behind it. In addition to that, we further reduce communication by skipping the
communication and update of corner cells, if they are not needed in a certain physics solver.
Optimizations include speeding up the flux calculation, avoid using split source terms which
need an extra ghost cells update by communication, and minimizing the conversion 
between primitive and conservative variables, (e.g., provide both primitive and conservative 
left and right state at cell interfaces after reconstruction for flux evaluation).

\begin{figure}
\includegraphics[width=\textwidth]{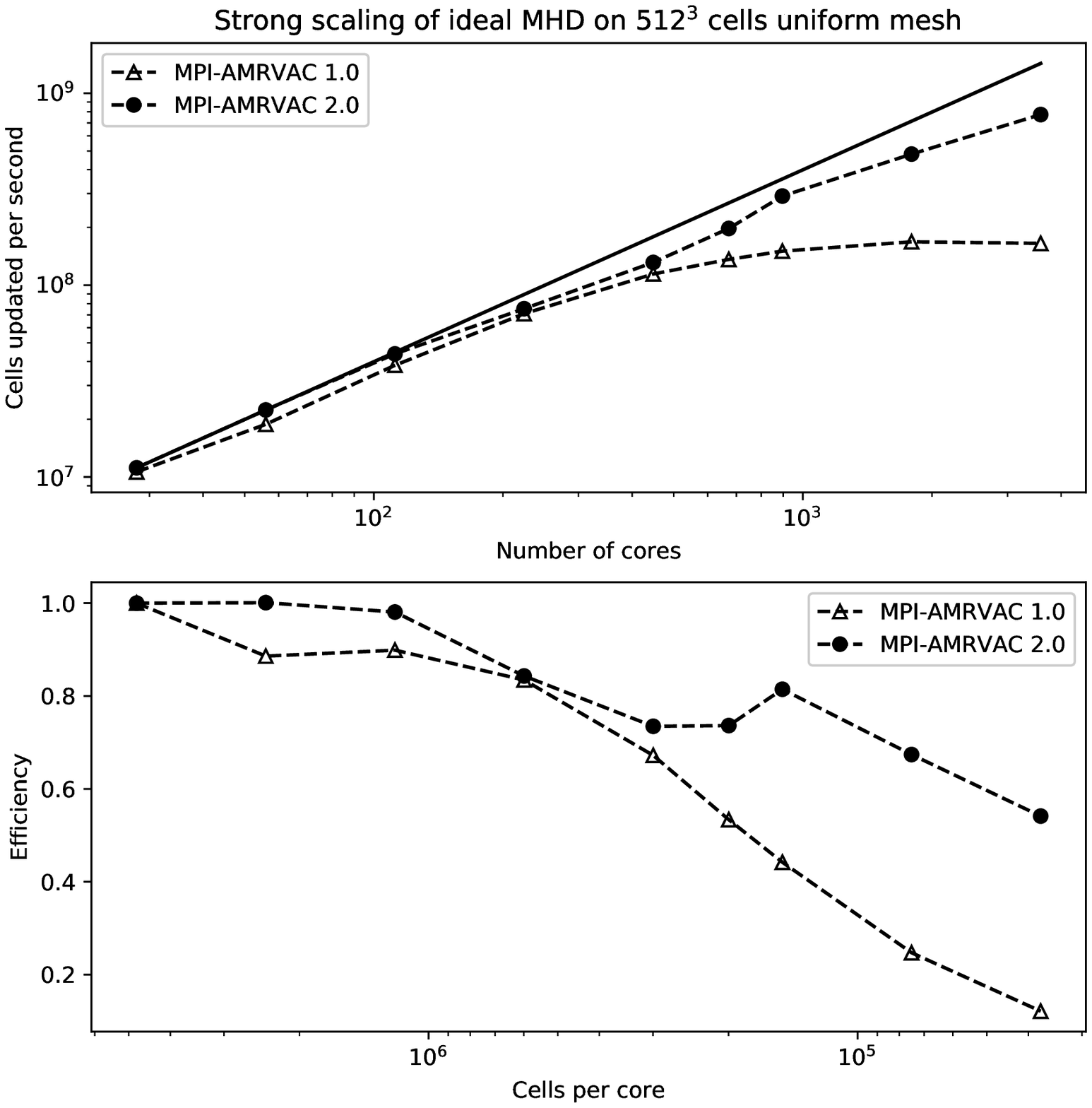}
\caption{Strong scaling of an ideal MHD application with $512^3$ cells uniform mesh for
\texttt{MPI-AMRVAC 2.0} shown in dots and for \texttt{MPI-AMRVAC 1.0} shown in empty trangles.
(a) Cells updated per second for increasing number of cores. The solid line shows the ideal scaling case. 
(b) Efficiency of the speed up, shown as function of the number of cells per core.
}
\label{fig:scalingimhd}
\end{figure}

\begin{figure}
\includegraphics[width=\textwidth]{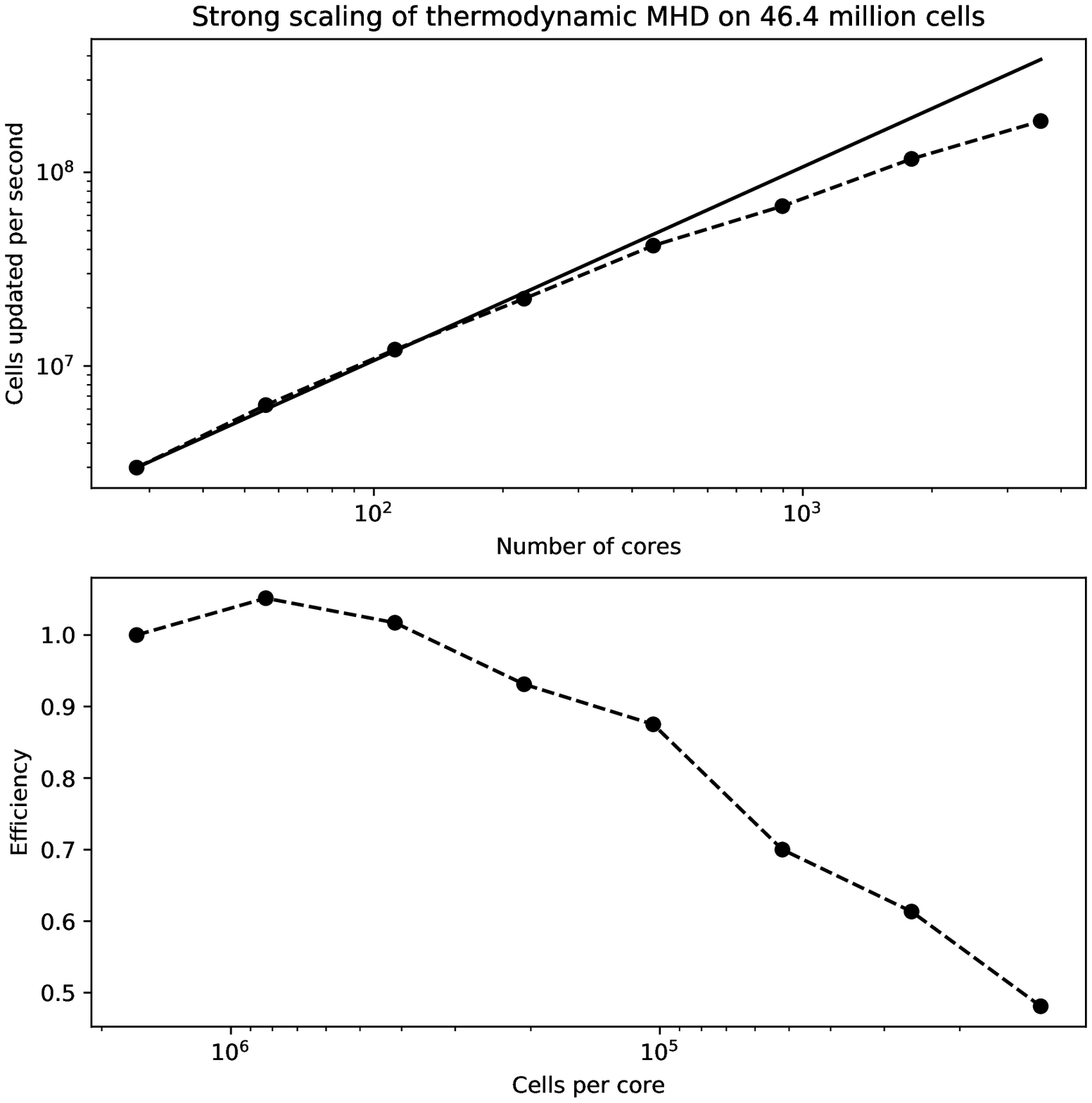}
\caption{Strong scaling of a thermodynamic AMR-MHD application with 46.4 million cells.
(a) Cells updated per second for increasing number of cores in dots connected with a
dashed line. The solid line shows the ideal scaling. (b) Efficiency of the speed up, shown as function of the number of cells per core.
}
\label{fig:scalingrmhd}
\end{figure}
With this new block ordering, we present two sets of tests to demonstrate strong 
scaling of \texttt{MPI-AMRVAC} on \texttt{Breniac}, the Flemish Tier-1 supercomputer. The
first one is solving a 3D blast wave problem in ideal MHD physics. The simulation box is
of size 2 by 2 by 2 on a uniform mesh of $512^3$ cells. The mesh is decomposed into 
$32^3$ blocks with each block of $16^3$ cells. Initially, gas pressure is 100 inside a central ball
with radius of 0.2 and 1 outside,  density is uniformly 1, magnetic field is uniform with 
$B_x=B_y=B_z=1$, and velocity is zero. We use the same schemes as explained in the beginning
of Section~\ref{sec:MFStest}, namely, HLL flux, \v{C}ada's compact third-order limiter, the
three-step Runge--Kutta time integration, and the diffusive approach to clean divergence of
magnetic field. To do strong scaling test, we run the same test with increasing number of 
cores, starting from 28 cores to 3584 cores. We plot speed as cells updated per second and
efficiency of speed up in Figure~\ref{fig:scalingimhd}, where the black line shows ideal scaling
with 100\% efficiency. The runs with the previous code version \texttt{MPI-AMRVAC} 1.0, plotted in empty triangle, 
stop speeding up after 1729 cores, while the new \texttt{MPI-AMRVAC} 2.0, plotted 
in dots, shows much better scaling. When running on 3584 cores (37449 cell per 
core), the test with \texttt{MPI-AMRVAC} 2.0 still has 54\% speed-up efficiency with respect to the 
speed of 28 cores. Note that each data point reported here is an average result of three repeated runs.

The second test comes from an actual solar application, where we have done a 3D coronal rain 
dynamics simulation \citep{Xia17}, which solves MHD coupled with parameterized heating, optically thin 
radiative cooling, and anisotropic thermal conduction as described in 
Section~\ref{sec:conduction}. We used an AMR mesh to resolve high-density, low-temperature 
coronal rain blobs when they form and fall down in a magnetic arcade~\citep{Xia17}. The 
AMR mesh is composed of $12^3$-cell (excluding ghost cells) blocks organized in 5 AMR levels. 
To do this strong scaling test, we take a snapshot (at time 46.3 min) from this coronal rain 
simulation, restart the simulation from that moment, and run for 10 time steps, with once
data output in the end. About 46.4 million cells in 26652 blocks are evolved. AMR mesh is 
reevaluated for (de)refinement every 3 time steps, although in the end the mesh structure 
and number of blocks does not change during the test. We use the same schemes as in the first 
test and the MFS technique mentioned in Section~\ref{sec:b0field}. The results of a strong scaling 
test are shown in Figure~\ref{fig:scalingrmhd}. More than 60\% efficiency is obtained when using up 
to about 2000 cores, with more than $5\times10^4$ cells per core.

\section{Generic framework improvements}
\label{sec:framework-improvements}

\texttt{MPI-AMRVAC} has been in development since early this century~\citep{Nool02}, 
and many scientists contributed code to it. These contributions usually
aimed to enable new research, for example by adding numerical methods or
extending physics models. In the most recent restructuring effort, we modernized 
the underlying framework, which will hopefully help in maintaining and developing the framework in the decades
to come. In this section, we describe improvements we made in this regard.

\subsection{Modernization of the framework}
\label{sec:modernization-framework}

\texttt{MPI-AMRVAC} is written in Fortran extended by the LASY syntax \citep{Toth97}.
After source files have been translated by a LASY preprocessor, they can be
compiled by a standard Fortran compiler. The main goal of the LASY syntax is to
be able to write dimension-independent code. However, the LASY syntax was also
used to for example:
\begin{itemize}
  \item (optionally) include other source files,
  \item perform vectorization (e.g., loop over components),
  \item define parameters for specific applications,
  \item turn parts of the code on or off.
\end{itemize}
This extended use of the preprocessor made it harder to understand the code,
especially as more and more features were added. The preprocessor in \texttt{MPI-AMRVAC 2.0}
has therefore been simplified; the only flags it now supports relate to dimensionality of 
the problem. Hence, the preprocessor is now only used to translate dimension-independent code to 1D, 2D or 3D. 
This simplification of the preprocessor also required another modernization: instead of 
source files that were optionally included as flags were turned on or off, we now make 
use of standard Fortran modules. This made it possible to compile the full framework into
a static library, including all the physics modules at once. This helps to ensure
code stays compatible (and compiles) whenever changes are made, and
significantly reduces compilation times. Furthermore, parallel compilation is
now supported.

The modifications imply that the user program now has to specify which type of major physics 
module to activate, for example the hydrodynamics or magnetohydrodynamics one (previously 
handled through the preprocessor and before compilation). As all modules are now available 
in the library version, this raises awareness of users to the full framework functionality, 
and each physics module now comes with its own settings and parameters, for example whether 
or not an energy equation should be used, or how many tracers (or dust species in coupled 
gas-dust hydro problems) to be included. Thanks to that, users can switch on/off special 
functionalities, like thermal conduction, radial stretching grid, particles etc., by just 
choosing appropriate parameters in an input parameter file without wasting time on 
recompilation, which was needed more frequently in the old code. Furthermore, we provide 
extensive support for customization to 
a large variety of applications by using function pointers. A user has to provide a
routine for setting initial conditions, but can also specify custom routines for
boundary conditions, additional source terms, custom output routines etc. 
There have also been many smaller changes to facilitate users, most importantly are those 
related to handling the coordinate system explicitly (by selecting e.g. Cartesian 2D or 
2.5D, or cylindrical, polar up to spherical 3D). Methods are encoded to automatically use 
the right number of ghost cells and the \texttt{MPI-AMRVAC 2.0} settings to run an 
actual application can now be specified in multiple parameter files.

\subsection{Automatic regression tests}
\label{sec:auto-tests}

A very important addition concerns regression tests for \texttt{MPI-AMRVAC}, in which their output is compared
to stored `correct' results. We intend to have a least one such test for each
type of major functionality that the code supports, varying dimensionality, equations to solve, or choice of algorithmic approach. The primary goal of these
tests is to ensure that code changes do not affect the result of existing
simulations. They can also help new users to explore \texttt{MPI-AMRVAC}'s functionality, and
to detect bugs and incompatibilities.

The implementation of these tests is as follows. First, a standard simulation is
scaled down so that it can be run in a short time, e.g. one to ten seconds on a
standard machine. We then specify the program to produce a log file of type
`regression\_test', which contains $\int {w}_i dV$ and $\int |{w}_i|^2 dV$
over time, where $\vec{w}$ is a vector of the $n_w$ (typically conservative) physical variables $w_i$ per cell, for all variables $i\in [1,n_w]$. By
comparing both the volume-integrated values of both cell-values and their squares,
errors in conservation properties and almost all errors in the shape of the
solution are detected. The resulting log file is then compared to a previously stored version, using a
finite error threshold in comparing values $a$ to $b$ as in:
$$|a - b| \leq \epsilon_\mathrm{abs} + \epsilon_\mathrm{rel}(|a| + |b|)/2,$$
where we by default use an absolute tolerance of
$\epsilon_\mathrm{abs} = 10^{-5}$ and relative tolerance of
$\epsilon_\mathrm{rel} = 10^{-8}$. These tolerances account for changes due to
numerical roundoff errors. Roundoff errors can for example occur when
computations are re-ordered (using different compilation flags and/or compilers)
or when a different number of processors is used. Most of the tests we have are
performed for several combinations of numerical schemes. They can be executed
automatically by running `make' in the \texttt{tests} folder of \texttt{MPI-AMRVAC}.

\subsection{Doxygen documentation}
\label{sec:doxygen-docs}

Having good documentation can increase the efficiency of both users and
maintainers. However, as it takes considerable effort to write such
documentation, this remains an ongoing process. For new contributions, we follow
these guidelines for inline documentation:
\begin{itemize}
  \item Provide a short description of a non-trivial function/subroutine,
  \item When a piece of code is hard to understand, briefly describe what it does,
  \item Variable names should be self-explanatory without getting lengthy.
\end{itemize}

For \texttt{MPI-AMRVAC 2.0} we make use of Doxygen (see \url{http://www.doxygen.org}), 
so that most documentation can be
written close to the code it documents. This helps when reading the code, but
also makes it simpler to update documentation when changing code. The
documentation is automatically generated every day and published on
\url{http://amrvac.org}, which provides a starting point for new users.

\subsection{New binary format}
\label{sec:binary-format}

Binary files are used to restart simulations, and they can also be
converted to formats supported by visualization toolkits. For \texttt{MPI-AMRVAC 2.0}, we
revised the binary format to make it future proof and simpler to use. The new
format is described in Table~\ref{tab:binary-header}. Below we briefly list some
of the changes:
\begin{itemize}
  \item The introduction of a version number to provide backwards functionality.
  \item The header is now at the start of the file, to simplify reading files.
  \item Byte offsets pointing to the location of the tree information and the
  block data allow new variables to be added at the end of the header.
  \item For each block, we also store its refinement, spatial index in the grid and the location of its data, to facilitate reading in data.
  \item Each grid block can now also store ghost cells.
\end{itemize}

\begin{table}
  \centering
  \begin{tabular}{c|c|c}
    \hline
    description & variable name & data type\\
    \hline
    \multicolumn{3}{c}{Header}\\
    \hline
    version number && int32\\
    offset tree info && int32\\
    offset block data && int32\\
    number of variables & nw & int32\\
    dimension of vectors & ndir & int32\\
    dimension of grid & ndim & int32\\
    maximum refinement level && int32\\
    number of grid leaves & nleafs & int32\\
    number of grid parents & nparents & int32\\
    time iteration count && int32\\
    simulation time && float64\\
    min. coordinates && float64(ndim)\\
    max. coordinates && float64(ndim)\\
    resolution of base-level mesh & domain\_nx& int32(ndim)\\
    resolution of a block & block\_nx & int32(ndim)\\
    geometry name && character(16)\\
    names of variables && character(nw*16)\\
    name of physics && character(16)\\
    number of physics parameters & nparams & int32\\
    parameters && float64(nparams)\\
    parameter names && character(nparams*16)\\
    \hline
    \multicolumn{3}{c}{Tree information}\\
    \hline
    leaf or parent && logical(nleafs+nparents)\\
    refinement level && int32(nleafs)\\
    spatial index && int32(ndim * nleafs)\\
    block offset && int64(nleafs)\\
    \hline
    \multicolumn{3}{c}{Block 1}\\
    \hline
    number of lower ghost cells & glo & int32(ndim)\\
    number of upper ghost cells & ghi & int32(ndim)\\
    block data && float64(1-glo:block\_nx+ghi, nw)\\
    \hline
    \multicolumn{3}{c}{Block 2}\\
    \hline
    \multicolumn{3}{c}{\ldots}\\
  \end{tabular}
  \caption{Description of the file format for binary files, in order of
    appearance in the file.}
  \label{tab:binary-header}
\end{table}

\section{Conclusions and outlook}

We provide an update on the open source software \texttt{MPI-AMRVAC}~\citep{Porth14}, 
suited for parallel, grid adaptive computations of hydrodynamic and 
MHD astrophysical applications. Motivated by recent applications 
in the context of transonic accretion flows onto compact objects, or trans-Alfv\'enic 
solar wind outflows, we combine grid stretching principles with block-adaptivity. Its 
advantage is demonstrated in 1D, 2D and 3D scenarios, on spherically symmetric Bondi 
accretion, steady planar Bondi--Hoyle--Lyttleton flows, 3D wind accretion in 
Supergiant X-ray binaries, and 2D solar wind. In especially solar physics oriented 
MHD applications, we show the benefits of a generalization of the original potential 
magnetic field splitting approach, which allows splitting off any magnetic field. 
This is demonstrated in tests with a 3D static force-free magnetic flux rope, in 
settings with a 2D magnetic null-point, and in resistive MHD scenarios about magnetic
reconnection in current sheets with either uniform or anomalous resistivity. 
The latter revisits a pioneering solar flare application with inclusion of anisotropic 
thermal conduction by \citet{Yokoyama01}, which can now be 
simulated at more realistic low plasma beta conditions. We provide details on our 
implementation for treating anisotropic thermal conduction in multi-dimensional MHD 
applications, generalizing slope limited symmetric schemes by 
\citet{Sharma07} from 2D to 3D configuration, and ring diffusion tests that
demonstrate its good accuracy and robustness in preventing from unphysical thermal 
flux from which all traditional schemes can not avoid. The new version of the code is 
shown to scale much better than the old one and its strong scaling is satisfactory up
to 2000 cores in AMR simulations on 3D solar coronal rain driven by thermal 
instabilities. Generic improvements are also reported: modularization and monolithic
compilation into a static library, the handling of automatic 
regression tests, the use of inline/online Doxygen documentation, and a new future-proof
 data format for input/output and restarts of simulations. 

The code is freely available on \url{http://amrvac.org}, and can be used for a fair 
variety of hydro to magnetohydrodynamic 
computations. In a forthcoming paper (Ripperda et al., in prep.), we will document 
the code capability to solve for particle dynamics using a variety of 
solvers for handling the governing Lorentz equation, or by using a set of 
ordinary differential equations for quantifying their gyro-averaged motion.  
This can be done in static, 
as well as in dynamic electro-magnetic fields obtained from MHD applications, 
as recently demonstrated in~\citet{ripp1,ripp2}. In combination with the 
capabilities documented in this paper, these improvements prepare for future 
applications of violent solar eruptive phenomena, and reconnection physics 
and particle acceleration associated with them. In the longer term, we may 
envision handling Poisson problems on the AMR hierarchy, or improving the 
treatment of radiative losses beyond the simplistic optically thin assumption. 

\acknowledgments
C.X., J.T., I.E.M., and E.C. want to thank FWO (Research Foundation Flanders) for the award of 
postdoctoral fellowship to them.
 This research was supported by FWO and by KU Leuven Project No. GOA/2015-014 and 
by the Interuniversity Attraction Poles Programme by the Belgian Science Policy Office (IAP P7/08 
CHARM). The simulations were conducted on the VSC (Flemish Supercomputer 
Center funded by Hercules foundation and Flemish government). 


\begin{thebibliography}{77}
\expandafter\ifx\csname natexlab\endcsname\relax\def\natexlab#1{#1}\fi

\bibitem[{Bisnovatyi-Kogan {et~al.}(1979)Bisnovatyi-Kogan, Kazhdan, Klypin,
  Lutskii, \& Shakura}]{Bisnovatyi-Kogan1979}
Bisnovatyi-Kogan, G.~S., Kazhdan, Y.~M., Klypin, A.~A., Lutskii, A.~E., \&
  Shakura, N.~I. 1979, Soviet Astronomy, 23

\bibitem[{Bondi(1952)}]{Bondi1952}
Bondi, H. 1952, \mnras, 112

\bibitem[{Bondi \& Hoyle(1944)}]{Bondi1944}
Bondi, H. \& Hoyle, F. 1944, \mnras, 104, 273

\bibitem[{Chan{\'e} {et~al.}(2005)Chan{\'e}, Jacobs, Van~der Holst, Poedts, \&
  Kimpe}]{Chane2005}
Chan{\'e}, E., Jacobs, C., Van~der Holst, B., Poedts, S., \& Kimpe, D. 2005,
  \aap, 432, 331

\bibitem[{Chan{\'e} {et~al.}(2008)Chan{\'e}, Poedts, \& Van~der
  Holst}]{Chane2008}
Chan{\'e}, E., Poedts, S., \& Van~der Holst, B. 2008, \aap, 492, L29

\bibitem[{Chen {et~al.}(2017)Chen, Frank, Blackman, Nordhaus, \&
  Carroll-Nellenback}]{Chen2017}
Chen, Z., Frank, A., Blackman, E.~G., Nordhaus, J., \& Carroll-Nellenback, J.
  2017, \mnras, 468, 4465

\bibitem[{{Cowie} \& {McKee}(1977)}]{Cowie77}
{Cowie}, L.~L. \& {McKee}, C.~F. 1977, \apj, 211, 135

\bibitem[{{Cunningham} {et~al.}(2009){Cunningham}, {Frank}, {Varni{\`e}re},
  {Mitran}, \& {Jones}}]{Cunninghametal2009}
{Cunningham}, A.~J., {Frank}, A., {Varni{\`e}re}, P., {Mitran}, S., \& {Jones},
  T.~W. 2009, \apjs, 182, 519

\bibitem[{Edgar(2004)}]{Edgar:2004ip}
Edgar, R.~G. 2004, New Astronomy Reviews, 48, 843

\bibitem[{{El Mellah}(2016)}]{ElMellah2016}
{El Mellah}, I. 2016, PhD thesis

\bibitem[{{El Mellah} \& Casse(2015)}]{ElMellah2015}
{El Mellah}, I. \& Casse, F. 2015, \mnras, 454, 2657

\bibitem[{{Fang} {et~al.}(2013){Fang}, {Xia}, \& {Keppens}}]{Fang13}
{Fang}, X., {Xia}, C., \& {Keppens}, R. 2013, \apjl, 771, L29

\bibitem[{{Fang} {et~al.}(2015){Fang}, {Xia}, {Keppens}, \& {Van
  Doorsselaere}}]{Fang15}
{Fang}, X., {Xia}, C., {Keppens}, R., \& {Van Doorsselaere}, T. 2015, \apj,
  807, 142

\bibitem[{{Feng} {et~al.}(2011){Feng}, {Zhang}, {Xiang}, {Yang}, {Jiang}, \&
  {Wu}}]{Feng11}
{Feng}, X., {Zhang}, S., {Xiang}, C., {Yang}, L., {Jiang}, C., \& {Wu}, S.~T.
  2011, \apj, 734, 50

\bibitem[{Foglizzo \& Ruffert(1996)}]{Foglizzo1996}
Foglizzo, T. \& Ruffert, M. 1996, \aap, 361, 22

\bibitem[{{Fromang} {et~al.}(2006){Fromang}, {Hennebelle}, \&
  {Teyssier}}]{Fromangetal2006}
{Fromang}, S., {Hennebelle}, P., \& {Teyssier}, R. 2006, \aap, 457, 371

\bibitem[{{Gombosi} {et~al.}(2002){Gombosi}, {T{\'o}th}, {De Zeeuw}, {Hansen},
  {Kabin}, \& {Powell}}]{Gombosi02}
{Gombosi}, T.~I., {T{\'o}th}, G., {De Zeeuw}, D.~L., {Hansen}, K.~C., {Kabin},
  K., \& {Powell}, K.~G. 2002, Journal of Computational Physics, 177, 176

\bibitem[{Groth {et~al.}(2000)Groth, De~Zeeuw, Gombosi, \& Powell}]{Groth2000}
Groth, C., De~Zeeuw, D.~L., Gombosi, T.~I., \& Powell, K.~G. 2000, Journal of
  Geophysical Research: Space Physics, 105, 25053

\bibitem[{{G{\"u}nter} {et~al.}(2005){G{\"u}nter}, {Yu}, {Kr{\"u}ger}, \&
  {Lackner}}]{Gunter05}
{G{\"u}nter}, S., {Yu}, Q., {Kr{\"u}ger}, J., \& {Lackner}, K. 2005, Journal of
  Computational Physics, 209, 354

\bibitem[{{Guo}(2015)}]{Guo15}
{Guo}, X. 2015, Journal of Computational Physics, 290, 352

\bibitem[{{Guo} {et~al.}(2016{\natexlab{a}}){Guo}, {Florinski}, \&
  {Wang}}]{Guo16}
{Guo}, X., {Florinski}, V., \& {Wang}, C. 2016{\natexlab{a}}, Journal of
  Computational Physics, 327, 543

\bibitem[{{Guo} {et~al.}(2016{\natexlab{b}}){Guo}, {Xia}, \&
  {Keppens}}]{Guo16b}
{Guo}, Y., {Xia}, C., \& {Keppens}, R. 2016{\natexlab{b}}, \apj, 828, 83

\bibitem[{{Guo} {et~al.}(2016{\natexlab{c}}){Guo}, {Xia}, {Keppens}, \&
  {Valori}}]{Guo16a}
{Guo}, Y., {Xia}, C., {Keppens}, R., \& {Valori}, G. 2016{\natexlab{c}}, \apj,
  828, 82

\bibitem[{{Harten} {et~al.}(1983){Harten}, {Lax}, \& {van Leer}}]{Harten83}
{Harten}, A., {Lax}, P.~D., \& {van Leer}, B. 1983, SIAM Review, 25, 35

\bibitem[{{Hendrix} \& {Keppens}(2014)}]{hendrix14}
{Hendrix}, T. \& {Keppens}, R. 2014, \aap, 562, A114

\bibitem[{{Hendrix} {et~al.}(2016){Hendrix}, {Keppens}, {van Marle}, {Camps},
  {Baes}, \& {Meliani}}]{hendrix16}
{Hendrix}, T., {Keppens}, R., {van Marle}, A.~J., {Camps}, P., {Baes}, M., \&
  {Meliani}, Z. 2016, \mnras, 460, 3975

\bibitem[{Hoyle \& Lyttleton(1939)}]{Hoyle:1939fl}
Hoyle, F. \& Lyttleton, R.~A. 1939, Mathematical Proceedings of the Cambridge
  Philosophical Society, 35, 405

\bibitem[{Jacobs {et~al.}(2005)Jacobs, Poedts, Van~der Holst, \&
  Chan{\'e}}]{Jacobs2005}
Jacobs, C., Poedts, S., Van~der Holst, B., \& Chan{\'e}, E. 2005, \aap, 430,
  1099

\bibitem[{{Keppens} {et~al.}(2008){Keppens}, {Meliani}, {van der Holst}, \&
  {Casse}}]{Keppens08}
{Keppens}, R., {Meliani}, Z., {van der Holst}, B., \& {Casse}, F. 2008, \aap,
  486, 663

\bibitem[{{Keppens} {et~al.}(2012){Keppens}, {Meliani}, {van Marle}, {Delmont},
  {Vlasis}, \& {van der Holst}}]{Keppens12}
{Keppens}, R., {Meliani}, Z., {van Marle}, A.~J., {Delmont}, P., {Vlasis}, A.,
  \& {van der Holst}, B. 2012, Journal of Computational Physics, 231, 718

\bibitem[{{Keppens} {et~al.}(2003){Keppens}, {Nool}, {T{\'o}th}, \&
  {Goedbloed}}]{Keppens03}
{Keppens}, R., {Nool}, M., {T{\'o}th}, G., \& {Goedbloed}, J.~P. 2003, Computer
  Physics Communications, 153, 317

\bibitem[{Keppens \& Porth(2014)}]{Keppens14}
Keppens, R. \& Porth, O. 2014, Journal of Computational and Applied
  Mathematics, 266, 87

\bibitem[{{Leroy} \& {Keppens}(2017)}]{Leroy17}
{Leroy}, M.~H.~J. \& {Keppens}, R. 2017, Physics of Plasmas, 24, 012906

\bibitem[{{Lohner}(1987)}]{Lohner87}
{Lohner}, R. 1987, Computer Methods in Applied Mechanics and Engineering, 61,
  323

\bibitem[{{Low}(1977)}]{Low77}
{Low}, B.~C. 1977, \apj, 212, 234

\bibitem[{{MacNeice} {et~al.}(2000){MacNeice}, {Olson}, {Mobarry}, {de
  Fainchtein}, \& {Packer}}]{MacNeice00}
{MacNeice}, P., {Olson}, K.~M., {Mobarry}, C., {de Fainchtein}, R., \&
  {Packer}, C. 2000, Computer Physics Communications, 126, 330

\bibitem[{{Manchester} {et~al.}(2004){Manchester}, {Gombosi}, {Roussev},
  {Ridley}, {de Zeeuw}, {Sokolov}, {Powell}, \& {T{\'o}th}}]{Manchester2004}
{Manchester}, W.~B., {Gombosi}, T.~I., {Roussev}, I., {Ridley}, A., {de Zeeuw},
  D.~L., {Sokolov}, I.~V., {Powell}, K.~G., \& {T{\'o}th}, G. 2004, Journal of
  Geophysical Research (Space Physics), 109, A02107

\bibitem[{{Mei} {et~al.}(2017){Mei}, {Keppens}, {Roussev}, \& {Lin}}]{Mei17}
{Mei}, Z.~X., {Keppens}, R., {Roussev}, I.~I., \& {Lin}, J. 2017, \aap, 604, L7

\bibitem[{{Meliani} \& {Keppens}(2010)}]{Meliani10}
{Meliani}, Z. \& {Keppens}, R. 2010, \aap, 520, L3

\bibitem[{{Meliani} {et~al.}(2007){Meliani}, {Keppens}, {Casse}, \&
  {Giannios}}]{Meliani07}
{Meliani}, Z., {Keppens}, R., {Casse}, F., \& {Giannios}, D. 2007, \mnras, 376,
  1189

\bibitem[{{Meyer} {et~al.}(2012){Meyer}, {Balsara}, \& {Aslam}}]{Meyer12}
{Meyer}, C.~D., {Balsara}, D.~S., \& {Aslam}, T.~D. 2012, \mnras, 422, 2102

\bibitem[{{Mignone} {et~al.}(2012){Mignone}, {Zanni}, {Tzeferacos}, {van
  Straalen}, {Colella}, \& {Bodo}}]{Mignone12}
{Mignone}, A., {Zanni}, C., {Tzeferacos}, P., {van Straalen}, B., {Colella},
  P., \& {Bodo}, G. 2012, \apjs, 198, 7

\bibitem[{{Monceau-Baroux} {et~al.}(2014){Monceau-Baroux}, {Porth}, {Meliani},
  \& {Keppens}}]{Monceau14}
{Monceau-Baroux}, R., {Porth}, O., {Meliani}, Z., \& {Keppens}, R. 2014, \aap,
  561, A30

\bibitem[{{Nool} \& {Keppens}(2002)}]{Nool02}
{Nool}, M. \& {Keppens}, R. 2002, Comp. Meth. Appl. Math., 2, 92

\bibitem[{{Parrish} \& {Stone}(2005)}]{Parrish05}
{Parrish}, I.~J. \& {Stone}, J.~M. 2005, \apj, 633, 334

\bibitem[{Phillips {et~al.}(1995)Phillips, Bame, Barnes, Barraclough, Feldman,
  Goldstein, Gosling, Hoogeveen, McComas, Neugebauer, {et~al.}}]{Phillips1995}
Phillips, J., Bame, S., Barnes, A., Barraclough, B., Feldman, W., Goldstein,
  B., Gosling, J., Hoogeveen, G., McComas, D., Neugebauer, M., {et~al.} 1995,
  Geophysical Research Letters, 22, 3301

\bibitem[{{Porth} {et~al.}(2014{\natexlab{a}}){Porth}, {Komissarov}, \&
  {Keppens}}]{Porth14RT}
{Porth}, O., {Komissarov}, S.~S., \& {Keppens}, R. 2014{\natexlab{a}}, \mnras,
  443, 547

\bibitem[{{Porth} {et~al.}(2014{\natexlab{b}}){Porth}, {Komissarov}, \&
  {Keppens}}]{Porth14MN}
---. 2014{\natexlab{b}}, \mnras, 438, 278

\bibitem[{{Porth} {et~al.}(2017){Porth}, {Olivares}, {Mizuno}, {Younsi},
  {Rezzolla}, {Moscibrodzka}, {Falcke}, \& {Kramer}}]{Porth17}
{Porth}, O., {Olivares}, H., {Mizuno}, Y., {Younsi}, Z., {Rezzolla}, L.,
  {Moscibrodzka}, M., {Falcke}, H., \& {Kramer}, M. 2017, Computational
  Astrophysics and Cosmology, 4, 1

\bibitem[{{Porth} {et~al.}(2014{\natexlab{c}}){Porth}, {Xia}, {Hendrix},
  {Moschou}, \& {Keppens}}]{Porth14}
{Porth}, O., {Xia}, C., {Hendrix}, T., {Moschou}, S.~P., \& {Keppens}, R.
  2014{\natexlab{c}}, \apjs, 214, 4

\bibitem[{{Powell} {et~al.}(1999){Powell}, {Roe}, {Linde}, {Gombosi}, \& {De
  Zeeuw}}]{Powell99}
{Powell}, K.~G., {Roe}, P.~L., {Linde}, T.~J., {Gombosi}, T.~I., \& {De Zeeuw},
  D.~L. 1999, Journal of Computational Physics, 154, 284

\bibitem[{Powell {et~al.}(1999)Powell, Roe, Linde, Gombosi, \&
  De~Zeeuw}]{Powell1999}
Powell, K.~G., Roe, P.~L., Linde, T.~J., Gombosi, T.~I., \& De~Zeeuw, D.~L.
  1999, Journal of Computational Physics, 154, 284

\bibitem[{{Ripperda} {et~al.}(2017{\natexlab{a}}){Ripperda}, {Porth}, {Xia}, \&
  {Keppens}}]{ripp1}
{Ripperda}, B., {Porth}, O., {Xia}, C., \& {Keppens}, R. 2017{\natexlab{a}},
  \mnras, 467, 3279

\bibitem[{{Ripperda} {et~al.}(2017{\natexlab{b}}){Ripperda}, {Porth}, {Xia}, \&
  {Keppens}}]{ripp2}
---. 2017{\natexlab{b}}, \mnras, 471, 3465

\bibitem[{{Rossmanith}(2004)}]{Rossmanith2004}
{Rossmanith}, J.~A. 2004, Computer Physics Communications, 164, 128

\bibitem[{Ruffert(1994)}]{Ruffert1994a}
Ruffert, M. 1994, \apj, 427, 342

\bibitem[{{Sharma} \& {Hammett}(2007)}]{Sharma07}
{Sharma}, P. \& {Hammett}, G.~W. 2007, Journal of Computational Physics, 227,
  123

\bibitem[{{Stone} {et~al.}(2008){Stone}, {Gardiner}, {Teuben}, {Hawley}, \&
  {Simon}}]{Stoneetal2008}
{Stone}, J.~M., {Gardiner}, T.~A., {Teuben}, P., {Hawley}, J.~F., \& {Simon},
  J.~B. 2008, \apjs, 178, 137

\bibitem[{{Tanaka}(1994)}]{Tanaka94}
{Tanaka}, T. 1994, Journal of Computational Physics, 111, 381

\bibitem[{{Toro} {et~al.}(1994){Toro}, {Spruce}, \& {Speares}}]{Toro94}
{Toro}, E.~F., {Spruce}, M., \& {Speares}, W. 1994, Shock Waves, 4, 25

\bibitem[{{T{\'o}th}(1997)}]{Toth97}
{T{\'o}th}, G. 1997, Journal of Computational Physics, 138, 981

\bibitem[{{T{\'o}th} {et~al.}(2008){T{\'o}th}, {Ma}, \& {Gombosi}}]{Toth08}
{T{\'o}th}, G., {Ma}, Y., \& {Gombosi}, T.~I. 2008, Journal of Computational
  Physics, 227, 6967

\bibitem[{{Tzeferacos} {et~al.}(2015){Tzeferacos}, {Fatenejad}, {Flocke},
  {Graziani}, {Gregori}, {Lamb}, {Lee}, {Meinecke}, {Scopatz}, \&
  {Weide}}]{Tzeferacosetal2015}
{Tzeferacos}, P., {Fatenejad}, M., {Flocke}, N., {Graziani}, C., {Gregori}, G.,
  {Lamb}, D.~Q., {Lee}, D., {Meinecke}, J., {Scopatz}, A., \& {Weide}, K. 2015,
  High Energy Density Physics, 17, 24

\bibitem[{{{\v C}ada} \& {Torrilhon}(2009)}]{Cada09}
{{\v C}ada}, M. \& {Torrilhon}, M. 2009, Journal of Computational Physics, 228,
  4118

\bibitem[{{van der Holst} \& {Keppens}(2007)}]{Holst07}
{van der Holst}, B. \& {Keppens}, R. 2007, Journal of Computational Physics,
  226, 925

\bibitem[{{van der Holst} {et~al.}(2008){van der Holst}, {Keppens}, \&
  {Meliani}}]{Holst08}
{van der Holst}, B., {Keppens}, R., \& {Meliani}, Z. 2008, Computer Physics
  Communications, 179, 617

\bibitem[{{van Leer}(1974)}]{vanLeer74}
{van Leer}, B. 1974, Journal of Computational Physics, 14, 361

\bibitem[{{Vlasis} {et~al.}(2011){Vlasis}, {van Eerten}, {Meliani}, \&
  {Keppens}}]{Vlasis11}
{Vlasis}, A., {van Eerten}, H.~J., {Meliani}, Z., \& {Keppens}, R. 2011,
  \mnras, 415, 279

\bibitem[{{Vreugdenhil} \& {Koren}(1993)}]{Koren93}
{Vreugdenhil}, C. \& {Koren}, B. 1993, Numerical methods for
  advection--diffusion problems (Braunschweig: Vieweg), 117--138

\bibitem[{{Wang} \& {Abel}(2009)}]{WangAbel2009}
{Wang}, P. \& {Abel}, T. 2009, \apj, 696, 96

\bibitem[{{Xia} {et~al.}(2012){Xia}, {Chen}, \& {Keppens}}]{Xia12}
{Xia}, C., {Chen}, P.~F., \& {Keppens}, R. 2012, \apjl, 748, L26

\bibitem[{{Xia} \& {Keppens}(2016)}]{Xia16}
{Xia}, C. \& {Keppens}, R. 2016, \apj, 823, 22

\bibitem[{{Xia} {et~al.}(2014){Xia}, {Keppens}, {Antolin}, \& {Porth}}]{Xia14}
{Xia}, C., {Keppens}, R., {Antolin}, P., \& {Porth}, O. 2014, \apjl, 792, L38

\bibitem[{{Xia} {et~al.}(2017){Xia}, {Keppens}, \& {Fang}}]{Xia17}
{Xia}, C., {Keppens}, R., \& {Fang}, X. 2017, \aap, 603, A42

\bibitem[{{Yokoyama} \& {Shibata}(2001)}]{Yokoyama01}
{Yokoyama}, T. \& {Shibata}, K. 2001, \apj, 549, 1160

\bibitem[{{Zhao} {et~al.}(2017){Zhao}, {Xia}, {Keppens}, \& {Gan}}]{ZhaoX17}
{Zhao}, X., {Xia}, C., {Keppens}, R., \& {Gan}, W. 2017, \apj, 841, 106

\bibitem[{{Ziegler}(2005)}]{Ziegler2005}
{Ziegler}, U. 2005, \aap, 435, 385

\end{thebibliography}

\end{document}